\documentclass{mpe_report}

\usepackage{psfig,graphicx,epsfig}
\usepackage{color}
\usepackage{amsmath,amssymb,epic,eepic,array}
\usepackage{natbib}

\begin{document}

\pagenumbering{arabic}
\setcounter{page}{44}

 \renewcommand{\FirstPageOfPaper }{ 44}\renewcommand{\LastPageOfPaper }{ 51}
\def\aj{AJ}%
\def\actaa{Acta Astron.}%
\def\araa{ARA\&A}%
\def\apj{ApJ}%
\def\apjl{ApJ}%
\def\apjs{ApJS}%
\def\ao{Appl.~Opt.}%
\def\apss{Ap\&SS}%
\def\aap{A\&A}%
\def\aapr{A\&A~Rev.}%
\def\aaps{A\&AS}%
\def\azh{AZh}%
\def\baas{BAAS}%
\def\bac{Bull. astr. Inst. Czechosl.}%
\def\caa{Chinese Astron. Astrophys.}%
\def\cjaa{Chinese J. Astron. Astrophys.}%
\def\icarus{Icarus}%
\def\jcap{J. Cosmology Astropart. Phys.}%
\def\jrasc{JRASC}%
\def\mnras{MNRAS}%
\def\memras{MmRAS}%
\def\na{New A}%
\def\nar{New A Rev.}%
\def\pasa{PASA}%
\def\pra{Phys.~Rev.~A}%
\def\prb{Phys.~Rev.~B}%
\def\prc{Phys.~Rev.~C}%
\def\prd{Phys.~Rev.~D}%
\def\pre{Phys.~Rev.~E}%
\def\prl{Phys.~Rev.~Lett.}%
\def\pasp{PASP}%
\def\pasj{PASJ}%
\def\qjras{QJRAS}%
\def\rmxaa{Rev. Mexicana Astron. Astrofis.}%
\def\skytel{S\&T}%
\def\solphys{Sol.~Phys.}%
\def\sovast{Soviet~Ast.}%
\def\ssr{Space~Sci.~Rev.}%
\def\zap{ZAp}%
\def\nat{Nature}%
\def\iaucirc{IAU~Circ.}%
\def\aplett{Astrophys.~Lett.}%
\def\apspr{Astrophys.~Space~Phys.~Res.}%
\def\bain{Bull.~Astron.~Inst.~Netherlands}%
\def\fcp{Fund.~Cosmic~Phys.}%
\def\gca{Geochim.~Cosmochim.~Acta}%
\def\grl{Geophys.~Res.~Lett.}%
\def\jcp{J.~Chem.~Phys.}%
\def\jgr{J.~Geophys.~Res.}%
\def\jqsrt{J.~Quant.~Spec.~Radiat.~Transf.}%
\def\memsai{Mem.~Soc.~Astron.~Italiana}%
\def\nphysa{Nucl.~Phys.~A}%
\def\physrep{Phys.~Rep.}%
\def\physscr{Phys.~Scr}%
\def\planss{Planet.~Space~Sci.}%
\def\procspie{Proc.~SPIE}%
\let\astap=\aap
\let\apjlett=\apjl
\let\apjsupp=\apjs
\let\applopt=\ao

\title{\textit{INTEGRAL} observations of PSR B0540-69}
\author{A. S\l{}owikowska\inst{1, 2} \and G. Kanbach\inst{2} \and
J. Borkowski\inst{1}  \and W. Becker\inst{2}}
\institute{Copernicus Astronomical Center, Rabia\'nska 8, 87-100 Toru\'n, Poland
\and
Max--Planck--Institut f\"ur extraterrestrische Physik, Giessenbachstra{\ss}e 1, 85740 Garching, Germany}
\maketitle

\section{Introduction}

\begin{table}[t]
\caption{Properties of PSR B0540-69 and PSR~B0531+21 (Crab pulsar).}
\label{Tab:comp}
\centering
\begin{tabular}{l c c}
\hline\hline
Parameter  & PSR B0540-69  & PSR B0531+21 \\
\hline
$P~(\rm{s})$                            &  0.05035 &  0.03308 \\
$\dot{P}~(\rm{s~s^{-1}})$   & $4.79 \times 10^{-13}$ & $4.23 \times 10^{-13}$\\
$n $                                           & 2.08 & 2.51 \\
$d~(\rm{kpc})$                        & 49.40  & 2.0 \\
$\tau_{c}~(\rm{kyr})$             & 1.67 & 1.24 \\
$B_{s}~(\rm{G})$                    & $4.97 \times 10^{12}$ & $3.78 \times 10^{12}$\\
$L_{sd}~(\rm{erg~s^{-1}})$ & $1.5 \times 10^{38}$ & $ 4.6 \times 10^{38}$  \\
Nebula size~(\rm{pc})             & $0.6 \times 0.9$ & $1.5 \times 1.5$ \\
\hline
\end{tabular}
\end{table}

PSR~B0540-69, a 50~ms pulsar located in the Large Magellanic Cloud (LMC), is one
of the most distant pulsars known ($d = 49.4\pm3.36~\rm kpc$, \citealt{Taylor1993}). 
\citet{Seward1984} discovered the pulsar at the position of supernova remnant SNR 0540-693 by using the \emph{Einstein} X-Ray Observatory. 
Shortly after its discovery a bright synchrotron wind nebula was confirmed in
the optical waveband \citep{Chanan1984}. The optical pulsations were soon detected by \citet*{Middleditch1985} with a mean pulsed magnitude of 22.5. In the radio band the pulsar is quite a faint source \citep{Manchester1993}, though \citet*{Johnston2003} recently reported the discovery of the first giant pulses from this pulsar. Since 1984, PSR~B0540-69 has been observed by: HST \citep{Boyd1995, Serafimovich2004},VLT \citep{Serafimovich2004}, BeppoSAX \citep{Mineo1999}, GINGA \citep{Deeter1999}, Chandra X-ray Observatory \citep{Gotthelf2000, Kaaret2001}, ASCA \citep{Hirayama2002} and RXTE \citep{dePlaa2003}.

PSR B0540-69 is often referred to as an extragalactic `twin' of the Crab pulsar (Tab.~\ref{Tab:comp}). Both pulsars have similar rotational parameters, characteristic age, spin-down luminosity and both are embedded in synchrotron plerionic nebulae. Here the similarity ends. Pulse profiles and spectra of PSR B0540-69 and PSR B0531+21 differ significantly. The Crab pulse profile shows a sharp double-peak structure, whereas the profile of LMC pulsar consists of a single broad peak. However, this broad pulse peak might be formed as a superposition of two Gaussian components separated of about 0.2 in phase \citep{dePlaa2003}. The X-ray spectrum of PSR B0540-60 differs from thet Crab spectrum \citep{dePlaa2003}. Additionally, unlike the Crab pulsar, PSR B0540-69 is not a $\gamma$-ray pulsar. 

\section{Observations and Data Analysis}
\begin{table*}
\caption{Time span of the \textit{INTEGRAL} LMC observations. A total of 813 ScWs within nine revolution were selected. First column gives the \emph{INTEGRAL} orbit (revolution) number. Second column shows the arbitrary numbers of ScWs  given in brackets, as well as a total number of ScWs per revolution. Next columns show revolution start and end time given in MJD TT and UTC, respectively.}
\label{Tab:idxfits}
\centering
\footnotesize
\begin{tabular}{l c c c c c}
\hline \hline
Rev & Number of ScWs & TSTART & TSTOP &
TSTART & TSTOP \\
\hline
0027 & (1-92)~~~92 & 52641.33 & 52643.98 & 
2003Jan02 at 07:57:54  & 2003Jan04 at 23:38:55 \\
0028 & (93-177)~~~85 & 52644.33 & 52646.97 & 
2003Jan05 at 08:01:21 & 2003Jan07 at 23:29:57 \\
0029 & (178-271)~~~94 & 52647.33 & 52649.97 &
2003Jan08 at 07:55:09  & 2003Jan10 at 23:20:02 \\
0033 & (272-367)~~~96 & 52659.29 & 52661.94 &
2003Jan20 at 06:58:11 & 2003Jan22 at 22:34:56 \\
0034 & (368-459)~~~92 & 52662.29 & 52664.93 &
2003Jan23 at 07:09:39 & 2003Jan25 at 22:19:59 \\
0035 & (460-552)~~~93 & 52665.27 & 52667.94 &
2003Jan26 at 06:29:29 & 2003Jan28 at 22:34:58 \\
0150 & (553-631)~~~79 & 53009.26 & 53011.93 &
2004Jan05 at 06:14:46 & 2004Jan07 at 22:29:59  \\
0151 & (632-726)~~~95 & 53012.23 & 53014.92 &
2004Jan08 at 05:37:33 & 2004Jan10 at 22:14:57 \\
0152 & (727-813)~~~87 & 53015.22 & 53017.93 &
2004Jan11 at 05:25:23 & 2004Jan13 at 22:21:52 \\
\hline
\end{tabular}
\end{table*}

The International Gamma-Ray Astrophysics Laboratory (\textit{INTEGRAL}) is a 15~keV -- 10~MeV gamma-ray observatory mission, consists of the spectrometer SPI and  the imager IBIS (Sec.~\ref{Sec:ibis}), with concurrent source monitoring in X-rays - 
the Joint European Monitor for X-rays (JEM-X, 3--35~keV, Sec.~\ref{Sec:jmx}) and in the optical range - the Optical Monitoring Camera (OMC, V, 500--600~nm).

We gathered more than one million seconds of PSR B0540-69 observations with \emph{INTEGRAL}. The data of one \textit{INTEGRAL} orbit, lasting about three sidereal days, are summarised as one revolution. The satellite was pointed to the LMC during nine revolutions (Tab.~\ref{Tab:idxfits}); six of them took place in January 2003 and three of them a year later, i.e.
in January 2004.  Because of the dithering of the \textit{INTEGRAL} satellite, single observations consist of many  \textit{pointings} that last about 30 minutes and are separated by \textit{slews}. Each pointing and slew, or only a part of them if they are too long, is called a \textit{Science Window} (ScW). Single observation usually consists of several ScWs.  Detailed information about the strategy of scientific observations and data
analysis can be found in the manuals published by \textit{INTEGRAL} Science Data Centre (ISDC) team\footnote{\texttt{http://isdc.unige.ch/index.cgi?Soft+download}}.
For the purpose of our project we used the \emph{INTEGRAL} standard software - `Off-line Science Analysis' (OSA, v5.1) to generate maps and source intensities. For the timing analysis we used our own IDL procedures.

\subsection*{Detailed description of the data selection}

\begin{table*}
\centering
\caption{Catalogue of selected sources used for the IBIS/ISGRI and JEM-X analysis. Coordinates are given in the increasing order of RA. Additionally the ISGRI (I\_FLAG) and JEM-X (J\_FLAG) detection flags are given; 0 - undetected, 1 - detected. However, the flag values are obtained for the standard settings of the data analysis performed by ISDC team. Especially the minimum detection sigma parameter could play a role here. The source coordinates were taken from the \textit{INTEGRAL} General Reference Catalogue (v.24) containing all sources detected by \textit{INTEGRAL} and being brighter than 1~mCrab above 1~keV. Only the pulsar position was changed. We used the coordinates given by \citet{Kaaret2001} on the basis of the Chandra observation.}
\label{Tab:usercat}
\begin{tabular}{c c c c c c}
\hline \hline
NAME & SOURCE\_ID & $\alpha~[\degr]$ & $\delta~[\degr]$ & I\_FLAG & J\_FLAG\\
\hline
LMC X-2    & J052029.2-715736 & 80.12167 & -71.96000 & 0 & 1 \\
LMC X-4    & J053249.2-662214 & 83.205     & -66.37056 & 1 & 1 \\
SN 1987A & J053530.0-691600 & 83.875     & -69.26667 & 0 & 0 \\
LMC X-3    & J053856.4-640501 & 84.735     & -64.08361 & 0 & 1 \\
LMC X-1    & J053938.7-694436 & 84.91125 & -69.74333 & 1 & 1 \\
PSR B0540-69 & J054007.7-692005 & 85.04675 &  -69.33194 & 1 & 0 \\
EXO 0748-676 & J074833.8-674509 & 117.1408 & -67.7525 & 1 & 0 \\
\hline
\end{tabular}
\end{table*}

We used all available \textit{INTEGRAL} data of PSR B0540-69. Until now the LMC was observed during the following nine revolutions: 0027, 0028, 0029, 0033, 0034, 0035, 0150, 0151 and 0152 (Tab.~\ref{Tab:idxfits}). It resulted in 813 ScWs  operated in the pointing mode. The exposure time per ScW is typically 2300 sec but occasionally ScWs as long as 9000 sec or as short as a few hundred seconds are encountered. The (arbitrary) ScW numbers corresponding to different revolutions are given in Tab.~\ref{Tab:idxfits}. Afterwards, we made the following selection. Using \textsf{idx\_find}\footnote{This procedure allows to select ScWs fulfilling terms of user criteria.} we requested for: ScW  being of pointing type, the target (pulsar) located inside the fully illuminated field of view (that is $9 \degr $ and $4 \fdg 8$ for IBIS and JEM-X, respectively), and both instrument modes being equal to 41, which indicates normal operations mode. We obtained two lists of ScWs fulfilling the terms of our criteria. The IBIS list consists of 617 ScWs, and the JEM-X one of 409 ScWs. None the less, in both cases the final number of used ScWs decreased during the analysis. It was caused by different reasons that are discussed in the according paragraphs.

Both, IBIS and JEM-X are coded mask instruments, therefore it is not possible to deal with one source at a time. Each source contributes to the background of the other sources detected in the field of view (FoV) of the instrument. Thus, all sources brighter or comparable with the target have to be included in the user (input) catalogue. For the purpose of the analysis we created our own catalogue. It is based on the general \textit{INTEGRAL} catalogue (v.24) and consists of seven sources visible in the IBIS FoV.  Detailed information of the selected sources is shown in Tab.~\ref{Tab:usercat}. The IBIS FoV is larger than the JEM-X FoV, therefore EX0~0748-676 was not visible by the X-ray instrument. Consequently only the first six sources listed in Tab.~\ref{Tab:usercat} were taken into account during the JEM-X analysis.

\section{IBIS/ISGRI data analysis and results}
\label{Sec:ibis}

\begin{table}
\caption{PSR~B0540-69 ephemeris.}
\label{Tab:eph_0540}
\centering
\begin{tabular}{l l}
\hline\hline
Parameter \& {\rm Value} & {\rm Value} \\
\hline
$\alpha_{2000}$  			&  			$05^{h} 40^{m} 11\fs221$ \\
$\delta_{2000}$   			&  			$-69\degr 19\arcmin 54\farcs98$ \\
 & \\
\multicolumn{2}{c}{Rev. 27, 28, 29, 33, 34 and 35}\\
\rm{Val. range (MJD)}            & 			52620--52670 \\
$t_0$ \rm{(TDB MJD)}            & 			52625.000000000 \\
$\nu_0$ \rm{(Hz)}                  & 			19.779298587915 \\
$\dot{\nu_0}~\rm{(10^{-10}~Hz~s^{-1})}$ 	&		$-1.87410$ \\   
$\ddot{\nu_0}~\rm{(10^{-20}~Hz~s^{-2})}$   & 		2.51 \\
 & \\
\multicolumn{2}{c}{Rev. 150, 151, 152}\\
\rm{Val. range (MJD)}  		&			53007--53062 \\
$t_0$ \rm{(TDB MJD)}		&			53007.000000000 \\
$\nu_0$ \rm{(Hz)}       		&			$19.773116586988(11189)$ \\
$\dot{\nu_0}~\rm{(10^{-10}~Hz~s^{-1})}$ & $-1.87256(11)$ \\
$\ddot{\nu_0}~\rm{(10^{-21}~Hz~s^{-2})}$  & $3.413(4.92)$\\
\hline
\end{tabular}
\end{table}

\begin{figure*}
\centerline{\hbox{
\psfig{file=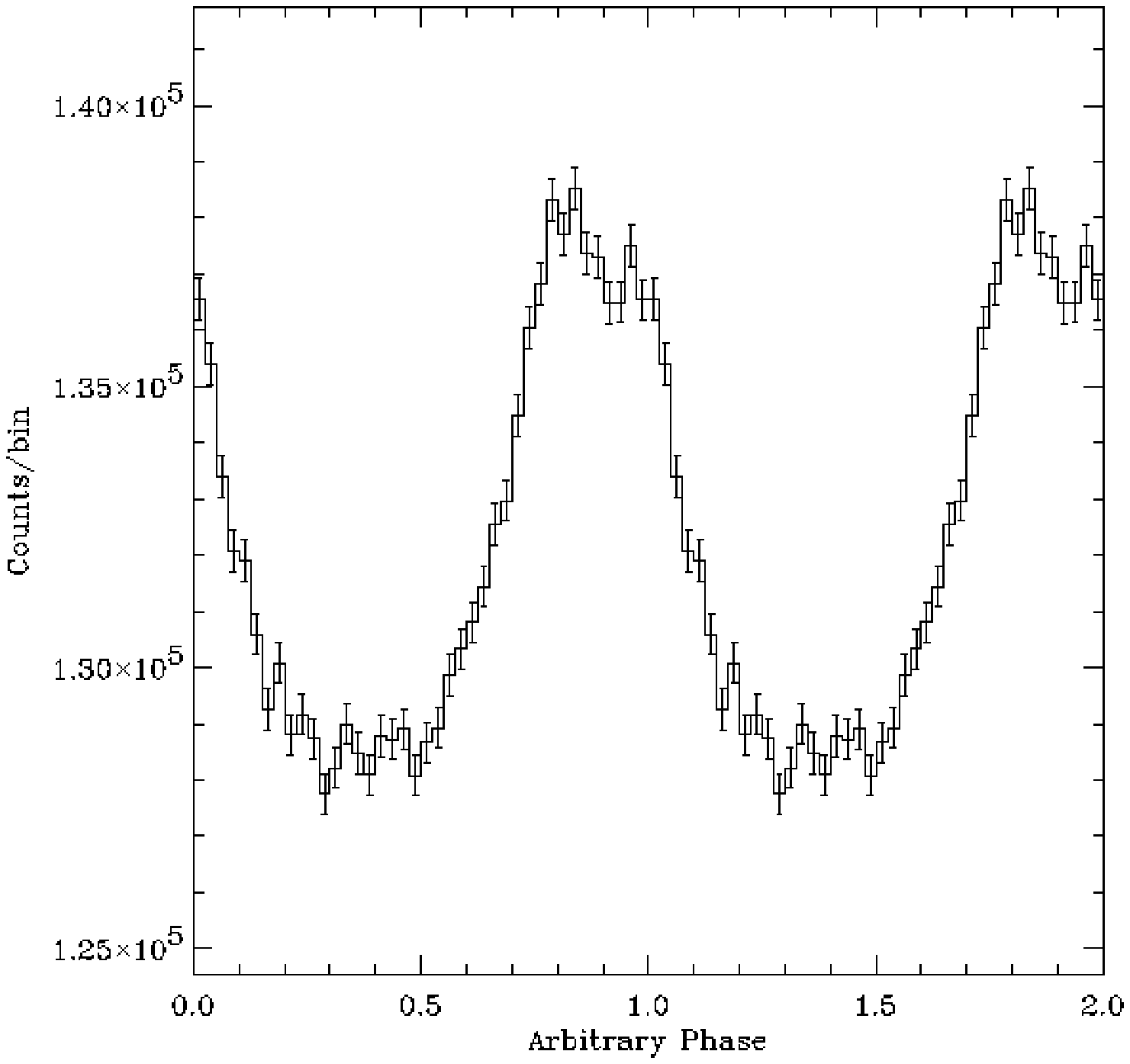,width=0.495\textwidth}
\psfig{file=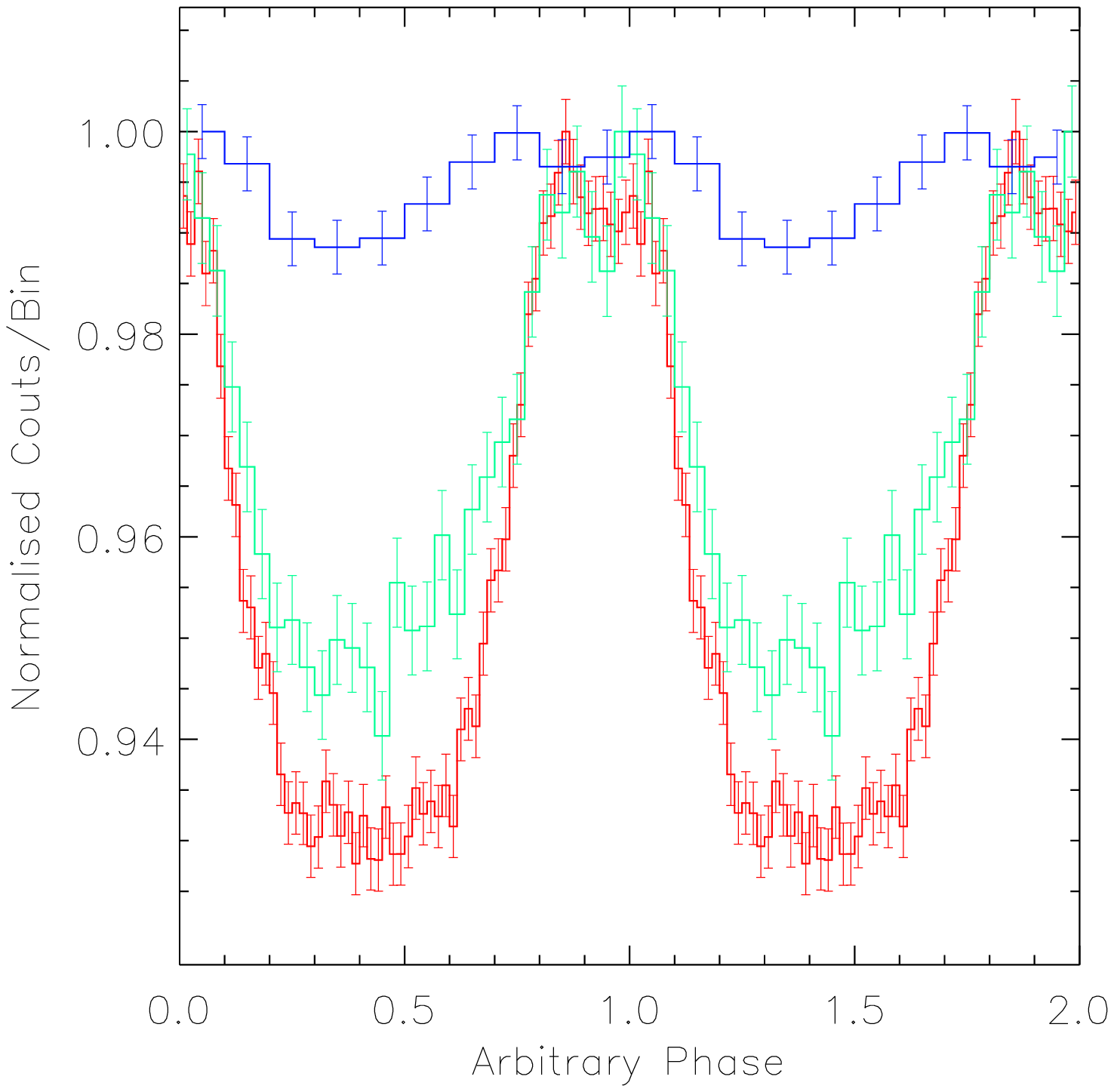,width=0.501\textwidth}
}}
\caption{\textit{Left}: RXTE~PCA~2--10~keV pulse profile of PSR~B0540-69 obtained form the data collected between December, 7 2002 and January, 30 2003. \textit{Right}: RXTE~PCA~ normalised pulse profiles of PSR~B0540-69 obtained form the data collected between January, 3 and February, 27 2004. The red (lowest), green (middle), and blue (highest) light curves corresponds to 2--10, 10--20, and 20--30~keV energy bands, respectively. The pulse duty cycle changes significantly with energy. The background level is unknown, therefore the pulsed fraction can not be measured. [Lucien Kuiper, private communication].
\label{rxte}}
\end{figure*}

\begin{figure*}[ht]
\centerline{\hbox{
\psfig{file=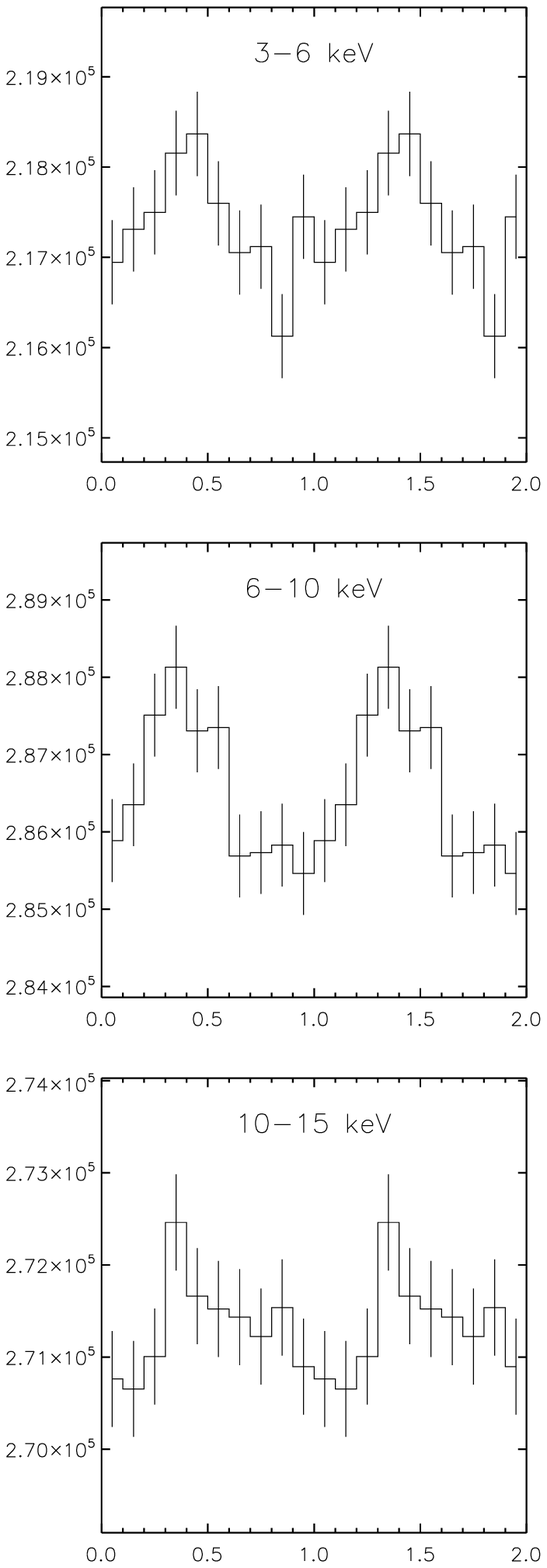,width=0.33\textwidth}
\psfig{file=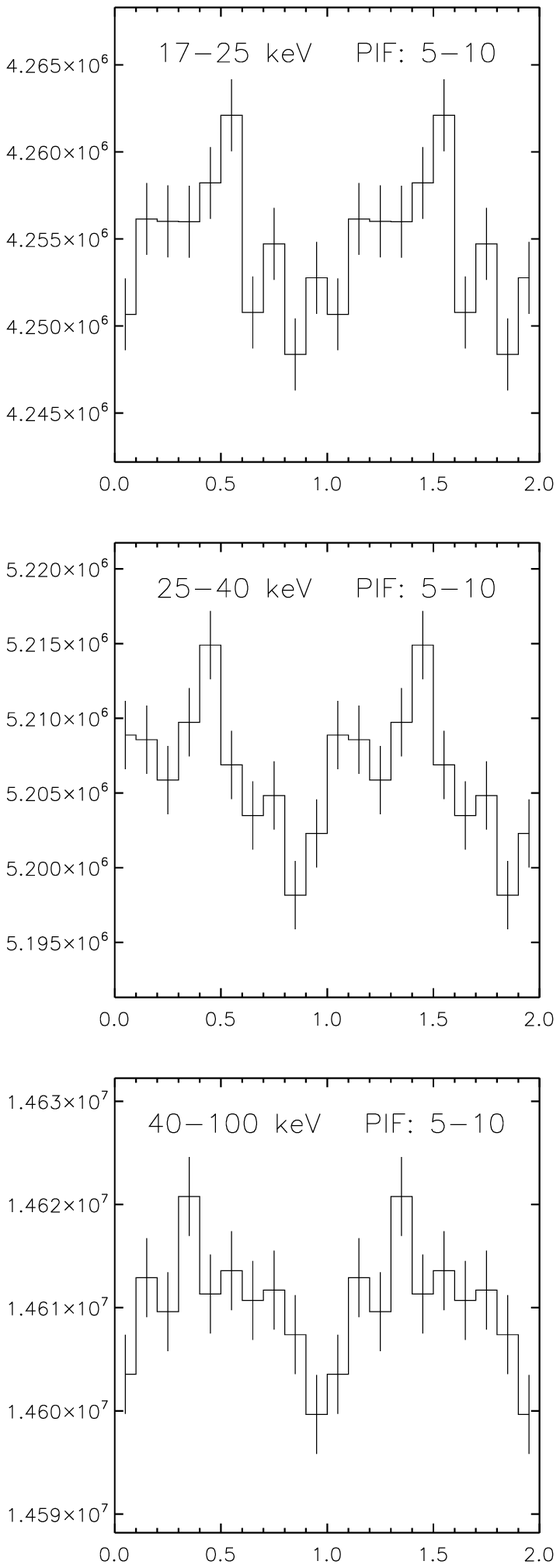,width=0.33\textwidth}
\psfig{file=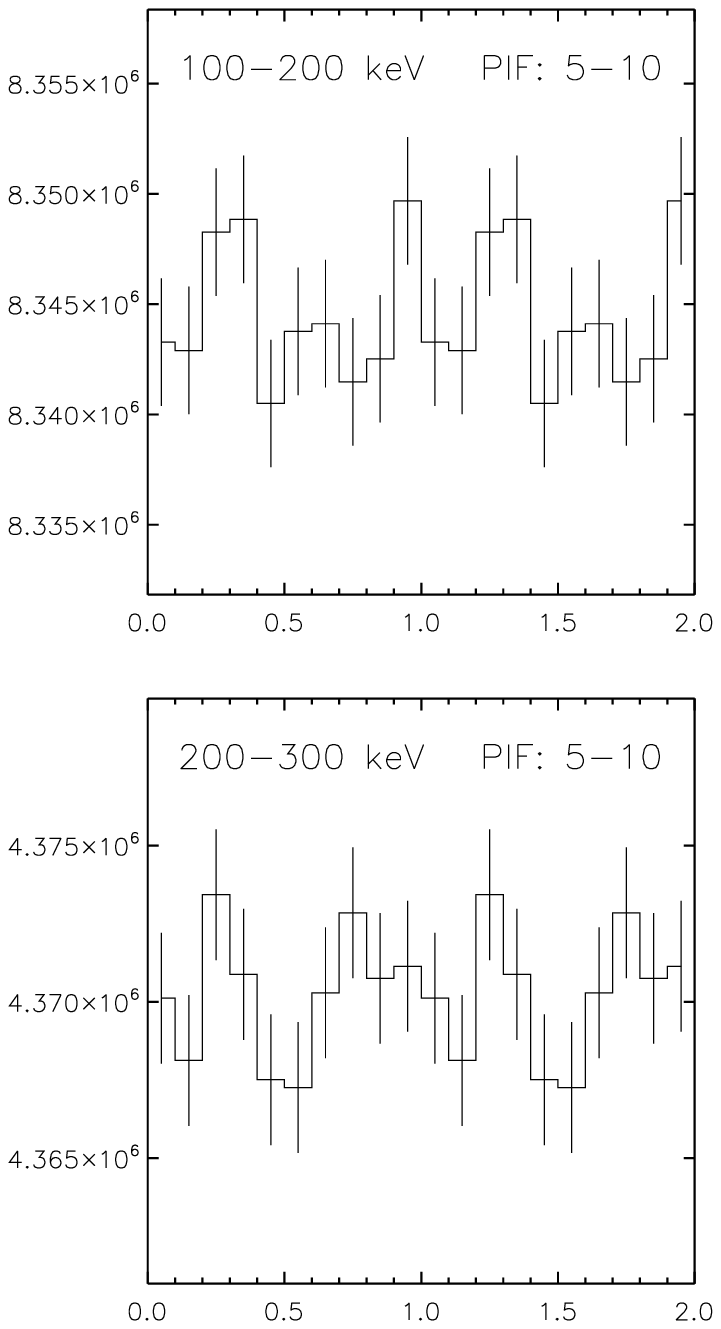,width=0.33\textwidth}
}}
\caption{PSR B0540-69 light curves obtained from JEM-X (\emph{left column}) and
IBIS (\emph{middle and right columns}) instruments. The energy bands are indicated in each panel, two cycles are shown for clarity. The PIF values in the plotted PIF ranges were multiplied by 10.
\label{Fig:lc_0540}}
\end{figure*}

IBIS is a gamma-ray telescope on-board of the \textit{INTEGRAL} satellite with powerful diagnostic capabilities of fine imaging, source identification and spectral sensitivity in both continuum and lines. It is able to localise weak sources at low energy to  better than a few arcminutes accuracy. Its energy resolution is 7\% at 100~keV and 9\% at 1~MeV. IBIS is a $\gamma$-ray imager consisting of two simultaneously operating detectors, ISGRI and PICsIT, covering the energy range from 20~keV to 10~MeV. The first layer (ISGRI) is made of Cadmium-Telluride (CdTe) solid-state detector and the second (PICsIT) of Cesium-Iodide (CsI) scintillator crystals. This configuration ensures a good broad line and continuum sensitivity over the wide spectral range covered by IBIS. Our analysis is based on the ISGRI single events, where the photons are stopped in one pixel of the ISGRI layer. The signal amplitude yields the energy of the incident photon. However, above 50~keV the energy is a function of not just the pulse height but also the pulse rise time (RISE\_TIME parameter, afterwords). In that case both  measured values are used to determine the energy of the photon. 

Lucien Kuiper (private communication) constructed pulsar ephemerides from observations with the Rossi X-ray Timing Explorer Mission  Proportional Counter Array (RXTE PCA). The RXTE data were taken between December, 7 2002 and January, 30 2003, as well as between January, 3 and February, 27 2004. The ephemerides obtained for both epochs are presented in Tab. \ref{Tab:eph_0540}, respectively, together with the INTEGRAL revolutions to which they are applicable. Note that the second frequency derivative for the time span 53007--53062 MJD (Rev. 150--152) is not significant. The RXTE~PCA~2--10~keV pulse profile constructed by using the data collected over December 2002 and January 2003 and folded accordingly to the pulsar ephemeris is shown in Fig. \ref{rxte}, \emph{left panel}. The pulse profiles obtained from the data collected between January  and February 2004 in three  energy bands: 2--10, 10--20, and 20--30~keV are shown in Fig.~\ref{rxte}, \emph{right panel}.
It is clear that the pulse duty cycle increases, while the pulsed fraction likely decreases with energy. The exact number of the pulsed fraction defined as $f = \frac{N_{P}}{N_{P}+N_{DC}}$ can not be derived because of the unknown background level. The ephemeris
for 2003 and 2004 observations are not absolute, and the alignment of the corresponding RXTE pulse profiles is not absolute either. Absolute timing would require defining the anchor point in the profile and referencing to this phase.

Since October 18, 2004 all public 	\textit{INTEGRAL} data are available in two formats: revision~1 (Rev.1) and revision~2 (Rev.2). In Rev.2, the correction of all JD time stamps for the offsets between the On Board Time (OBT) of each instrument is done. In Rev.2 the data correction (COR) step, as well as the instrumental Good Time Interval (GTI) and deadline handling (DEAD) steps are performed by the ISDC using OSA (v.4.2) at the science window level. During the COR step the data are corrected for instrumental effects, such as energy and position corrections, while the GTI step generates, selects and merges Good Time Intervals to produce a unique GTI that is later used for selecting events. The net observing time is also computed in this step. Dead and live times are computed during the DEAD step of the analysis. The dead time is the time during which the instrument was not capable (for different reasons) to register the incoming photons  within a GTI. However, the data correction implemented in OSA 5.1 is much better. Therefore, it is highly recommended by the ISDC team to rerun these three steps (COR, GTI, DEAD) over the used data. We performed these correction steps over all selected ScWs, which was a very computer intensive task.

It is known from the ISGRI limitations that a problem on-board IBIS causes, under some circumstances, event times to be shifted by 2 seconds. The OSA software corrects for it, but still the possible jumps might occur. Such behaviour can be verified by checking the FITS flag keyword TIMECORR. For accurate timing it is recommended to use only the data with no possible jump, i.e. TIMECORR=0. Therefore, we used only ScWs with TIMECORR equal 0. The most up to date version of the known limitations, including list of ScWs for which the  two-second jumps have been detected, can be found on the ISDC web pages \footnote{ \texttt{http://isdc.unige.ch/Soft/download/osa/osa\_sw/\\osa\_sw-5.1/osa\_issues.txt}}.

A standard and stand-alone tool  \textsf{ii\_pif\_built} is used to calculate the ISGRI Photon Illumination Fraction (PIF) for a given source. Once the PIF map is made, it is straightforward to filter event lists based on the PIF threshold. By applying the PIF filter, one can reduce the background and optimise the signal to noise ratio. From experience, we choose PIF$ > 0.5$ as the event selection criterion. We run the \textsf{ii\_pif\_built} procedure with pulsar position ($\alpha_{2000} = 85 \fdg 046754167$, $\delta_{2000} = -69 \fdg 331938889$, \citet{Kaaret2001}) as an input parameter to compute the PIF for each event. By using the \textsf{evts\_extract} script only single events (evttype=0, $7 < \rm{RISE\_TIME} < 90$) were selected. Additionally, the noisy pixels (SELECT\_FLAG=0) were filtered out. The procedure \textsf{evts\_extract} allows to barycentre the events, however at this point we did not use it. Firstly,  the instrument time corrections were applied, i.e. $111\pm10\rm \mu s$ and $185\pm10\rm \mu s$ for IBIS and JEM-X, respectively \citep{Walter2003} and then derived times of arrivals (TOAs) were barycentred. Afterwards, only events with the energy $\leq 300$~keV were selected. For these events we stored their TOAs, PIF and energy values. TOAs belonging to the first and second set of observations were then folded according to the corresponding RXTE ephemeris (Tab. \ref{Tab:eph_0540}). Hence, we obtained two sets of the pulsar light curves in five energy bands: 17--25, 25--40, 40-100, 100-200, and 200-300~keV. The ephemerides are not absolute, therefore to sum up the light curves from both observations we used a cross-correlation method to determine their relative phase offset. As the template we used the light curve with the highest significance, i.e. the light curve obtained for JEM-X data in the 6-10~keV energy band (see Sec.~\ref{Sec:jmx}). In the light curves only the events with $\rm{PIF} > 0.5$ are included, i.e. with probability of illuminating the pixel by the source being higher than 50\%. The results are presented in Fig.~\ref{Fig:lc_0540}, \emph{middle and right columns}.

\begin{figure}
\centerline{\psfig{file=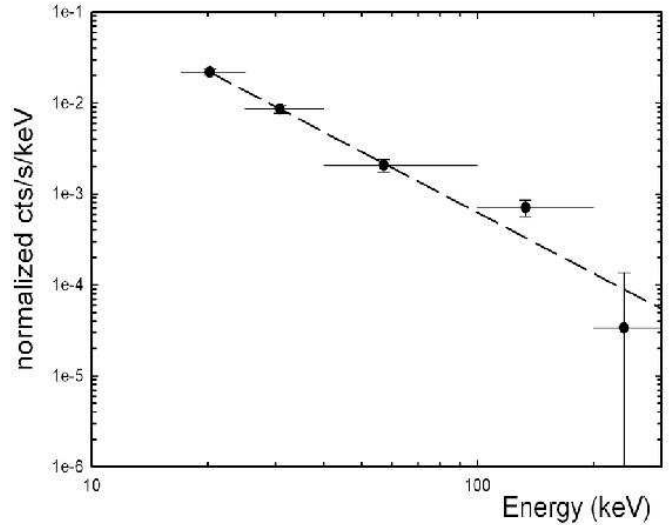,width=0.50\textwidth}}
\caption{PSR B0540-69 spectrum  in 17--300~keV energy range. Above 200~keV only upper limit was derived. The dashed line indicates the power law fit.
\label{Fig:spec_0540}}
\end{figure}

\begin{figure*}
\centerline{\hbox{
\psfig{file=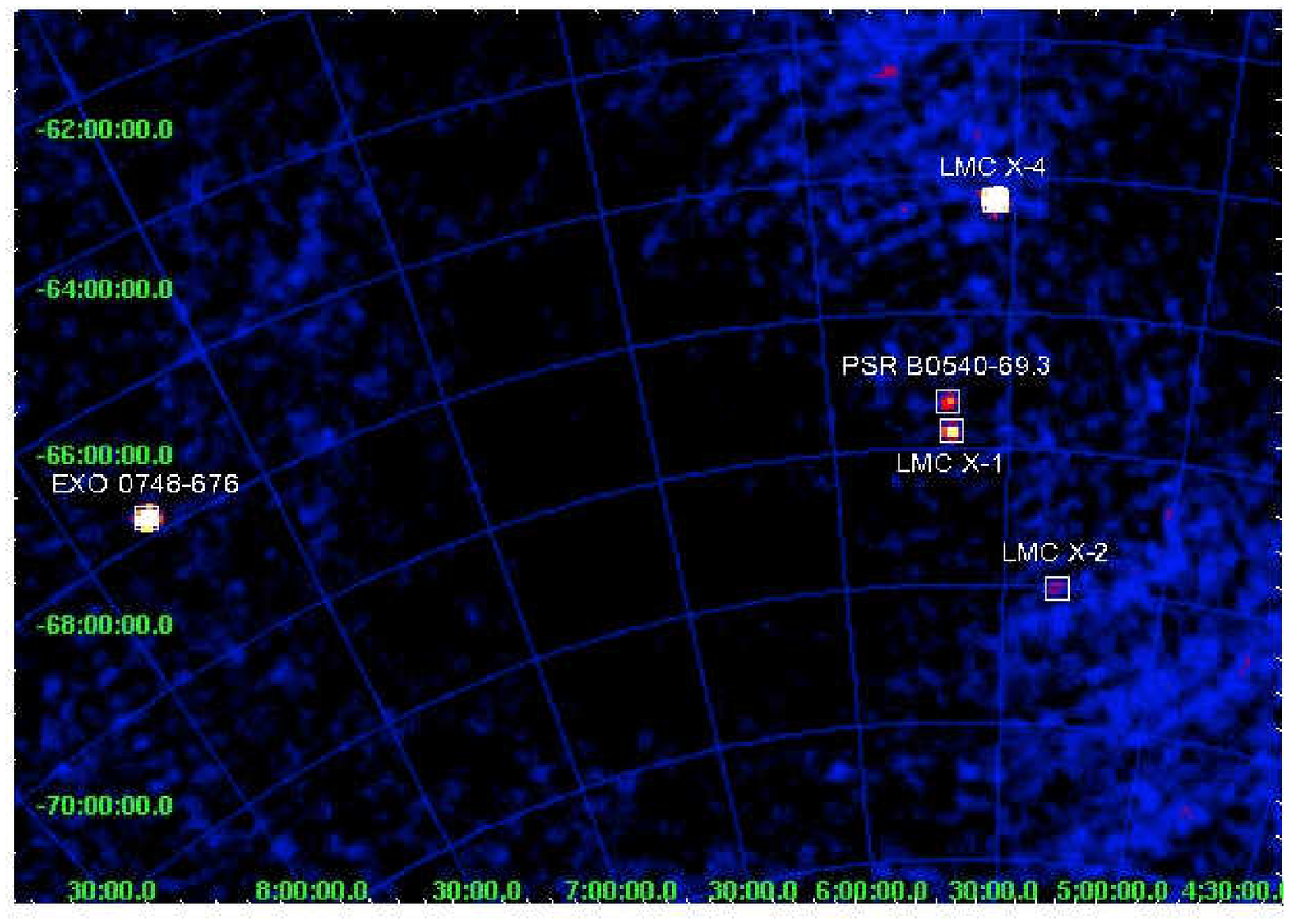,width=0.45\textwidth}
\psfig{file=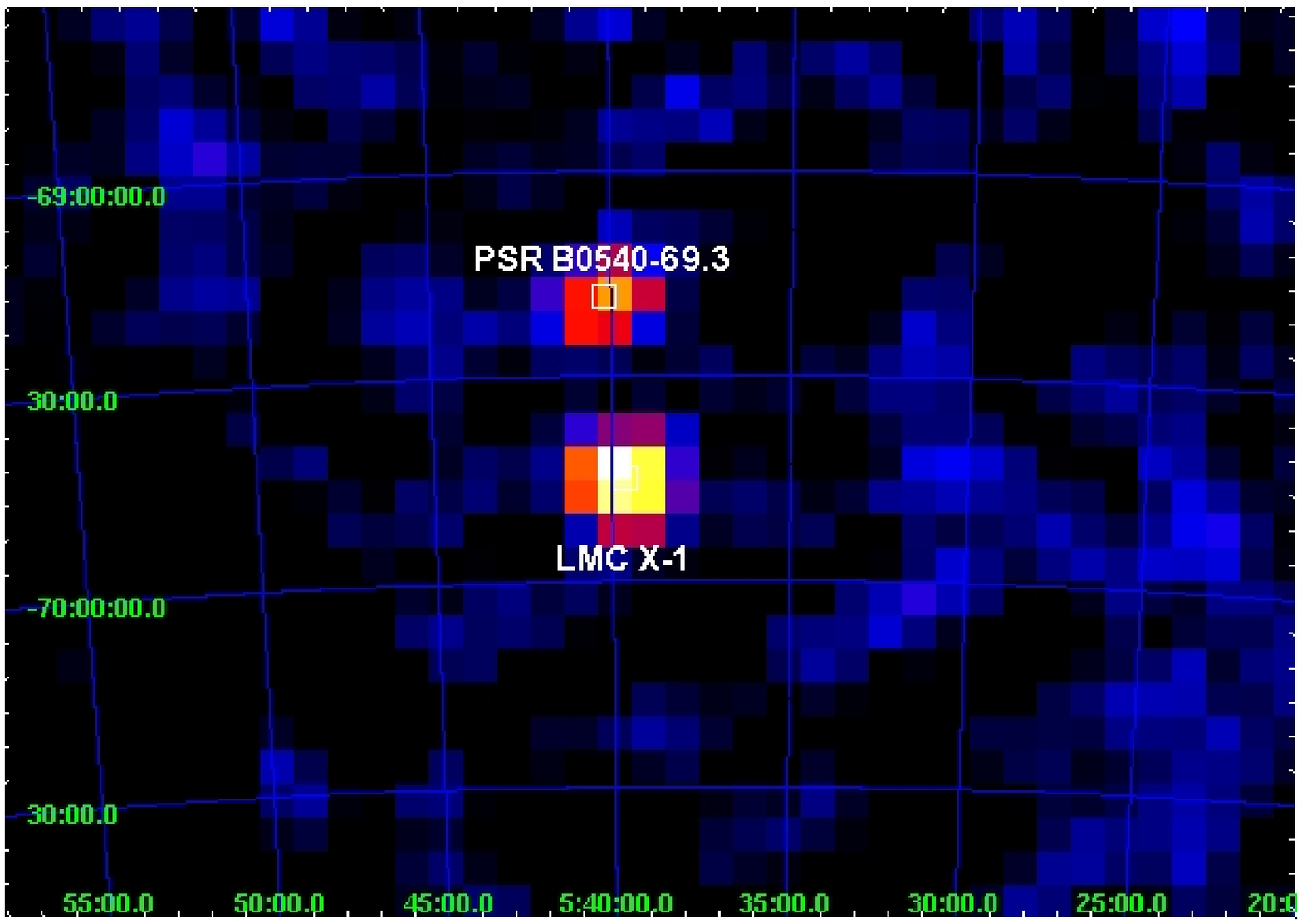,width=0.45\textwidth}
}}
\centerline{\hbox{
\psfig{file=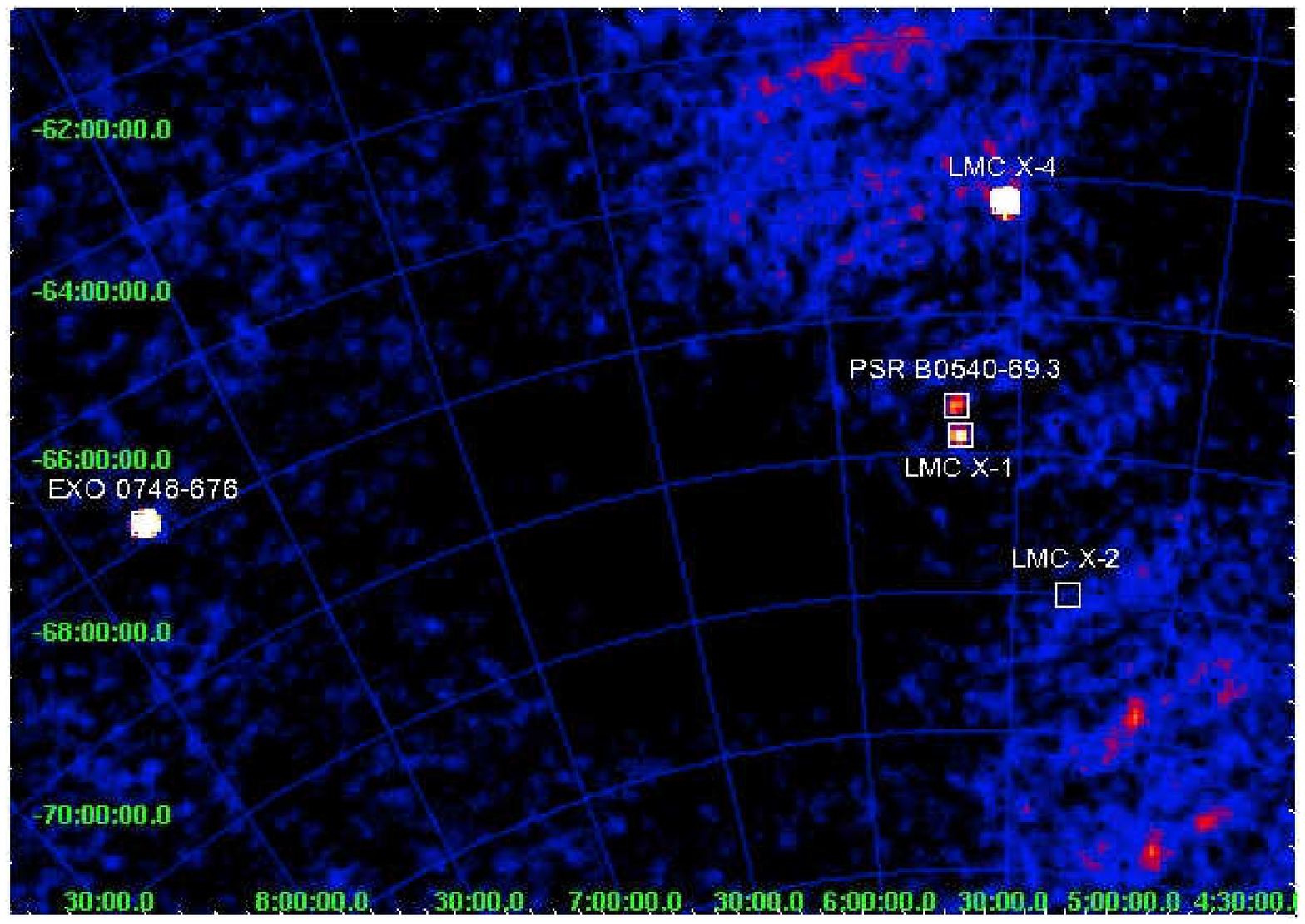,width=0.45\textwidth}
\psfig{file=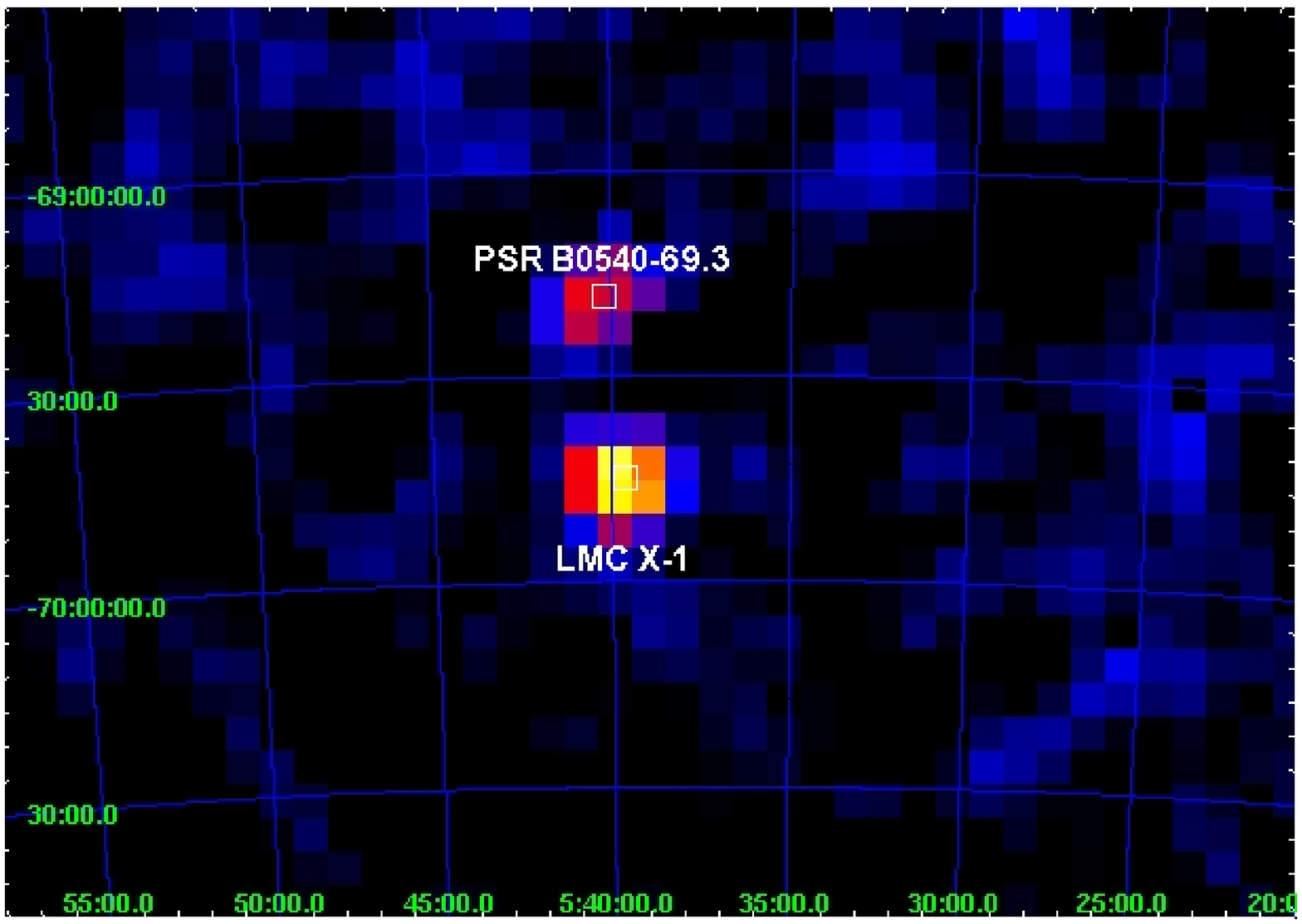,width=0.45\textwidth}
}}
\centerline{\hbox{
\psfig{file=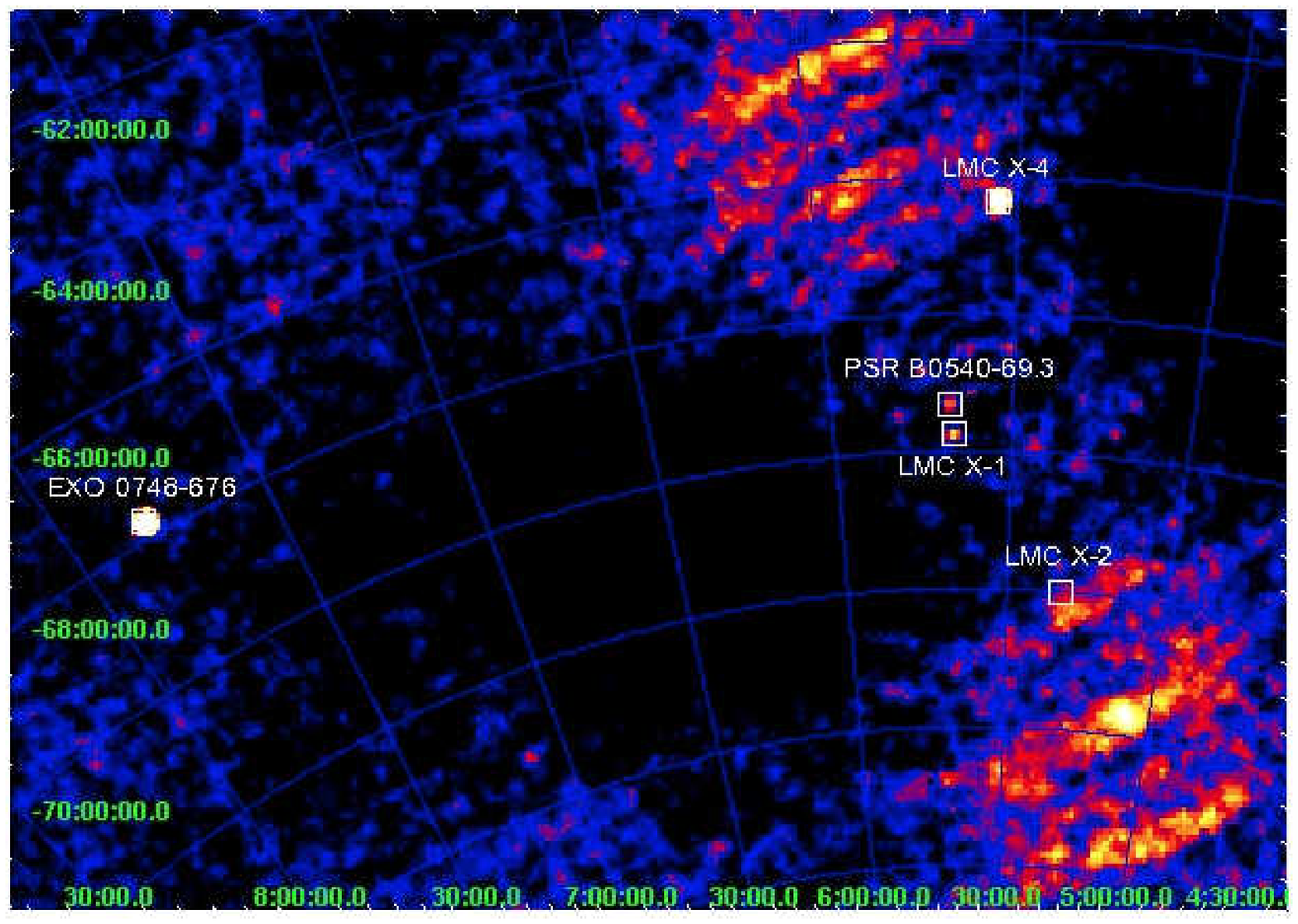,width=0.45\textwidth}
\psfig{file=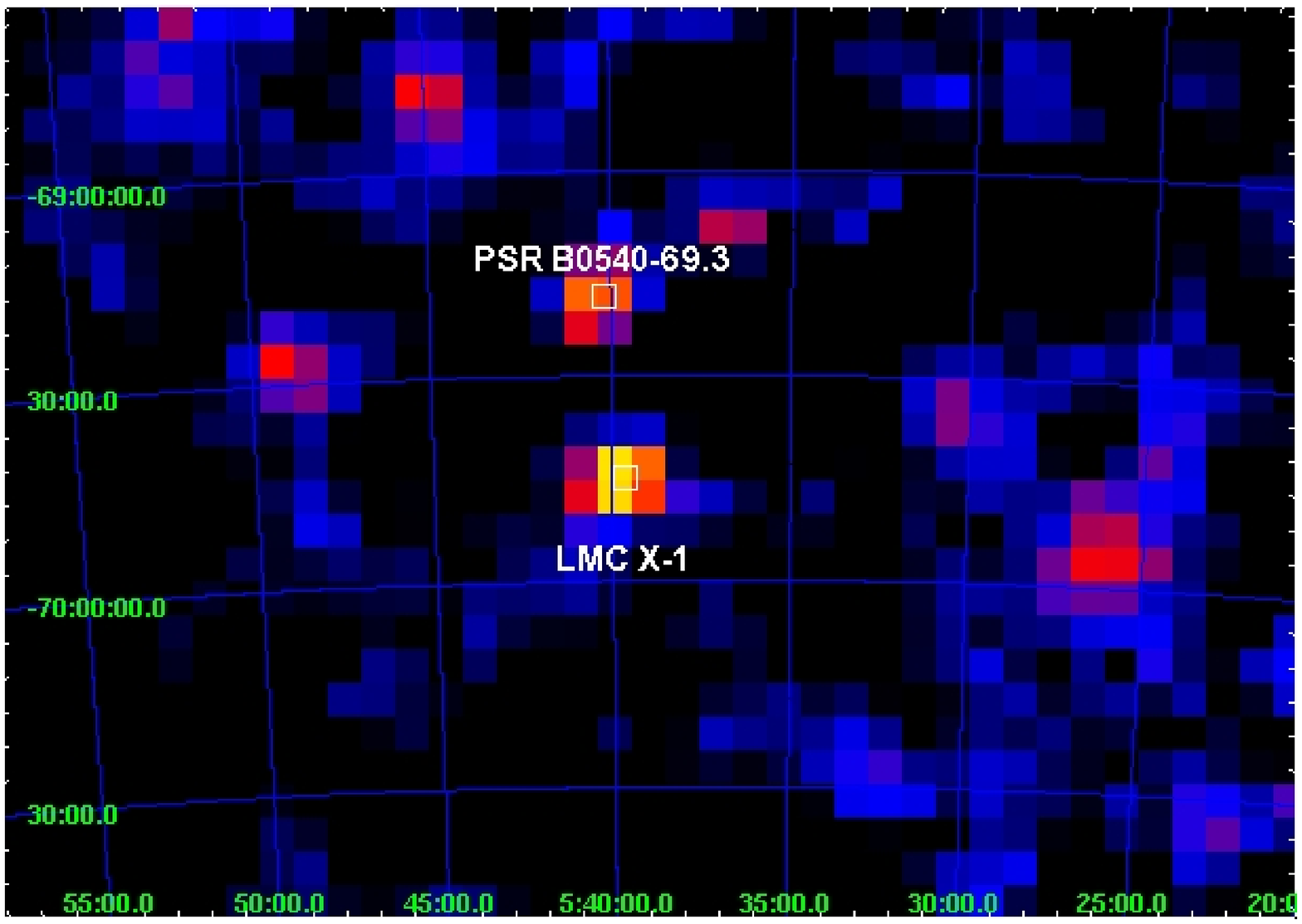,width=0.45\textwidth}
}}
\centerline{\hbox{
\psfig{file=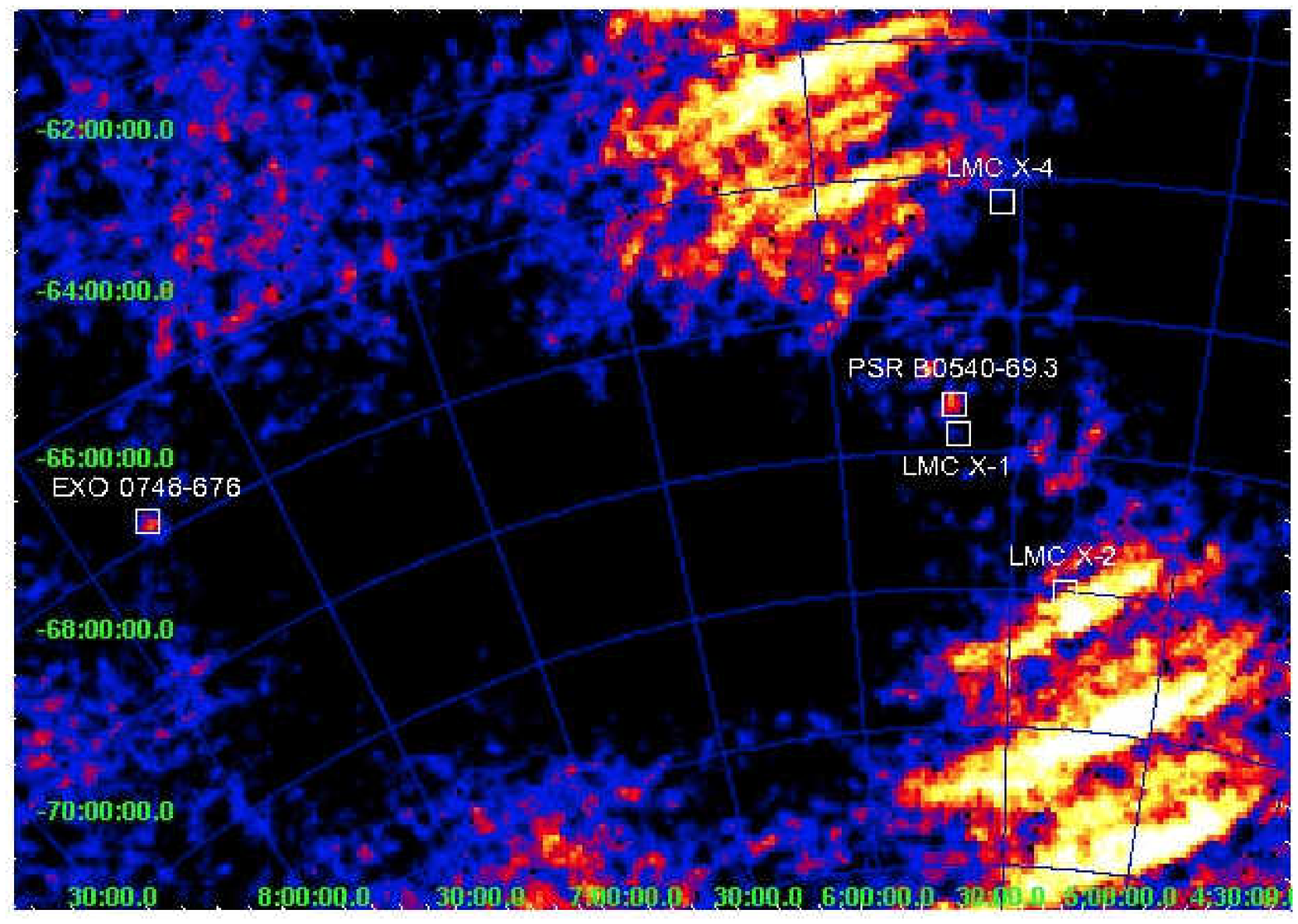,width=0.45\textwidth}
\psfig{file=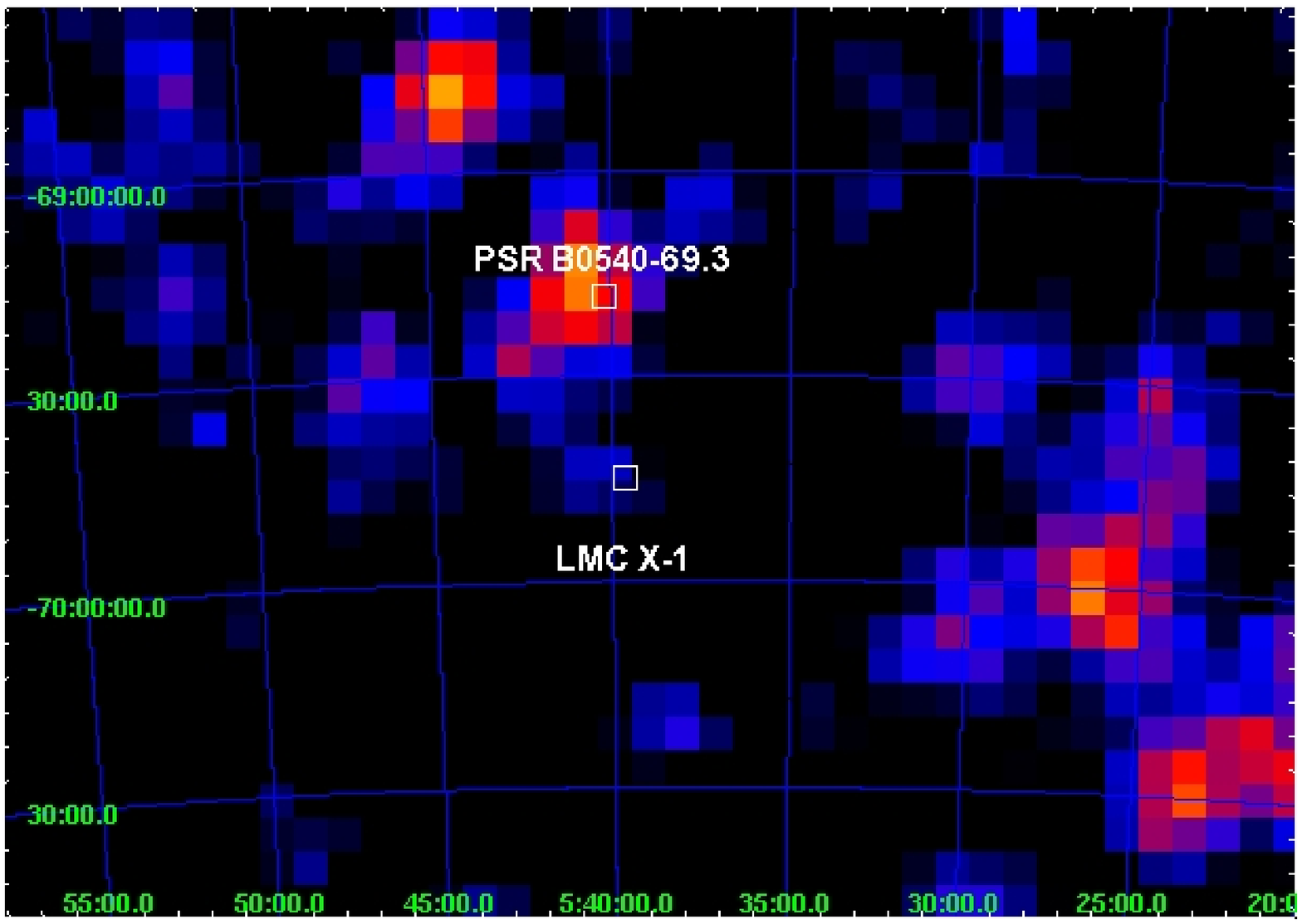,width=0.45\textwidth}
}}
\caption{The IBIS/ISGRI significance mosaic images of the LMC region (\emph{left}) and the close-up of the pulsar neighbourhood (\emph{right}). Maps are for the 17--25, 25--40, 40--100 and 100--200~keV energy ranges from top to bottom, respectively.  
PSR B0540-69 was detected with following significance 13.8$\sigma$, 11.2$\sigma$,
7.6$\sigma$ and 6.2$\sigma$, respectively.
\label{Fig:isgri_mosa2}}
\end{figure*}

By standard data processing (\textsf{ibis\_science\_analysis}) we produced the sky images for each individual ScW. The source is very faint, therefore it was not detected with high significance levels in single ScWs. The OSA software allows to gather the single ScWs belonging to different observations. Thus, we used all selected ScWs to produce the mosaic maps. They correspond to the same five energy bands used for the light curve extraction (Fig.~\ref{Fig:lc_0540}). In the mosaic maps the PSR B0540-69 was detected in the four lower energy bands with the following significances: 13.8$\sigma$, 11.2$\sigma$, 7.6$\sigma$ and 6.2$\sigma$, respectively (Fig.~\ref{Fig:isgri_mosa2}). There was no significant detection above 200~keV.

For the faint sources the recommended spectral extraction method is based on the mosaic sky images. To derive the spectrum we used the count rate values obtained from these images. We derive the spectrum of the total source exclusively from the ISGRI data, treating the calibration of JEM-X as not applicable to such a weak source. The photon spectrum can be fitted with a power law of index -2.22, which is compatible with the result found by \citet{Gotz2006}. The spectrum and the power law fit are shown in Fig.~\ref{Fig:spec_0540}.

\section{JEM-X data analysis and results}
\label{Sec:jmx}

\begin{figure*}
\centerline{\hbox{
\psfig{file=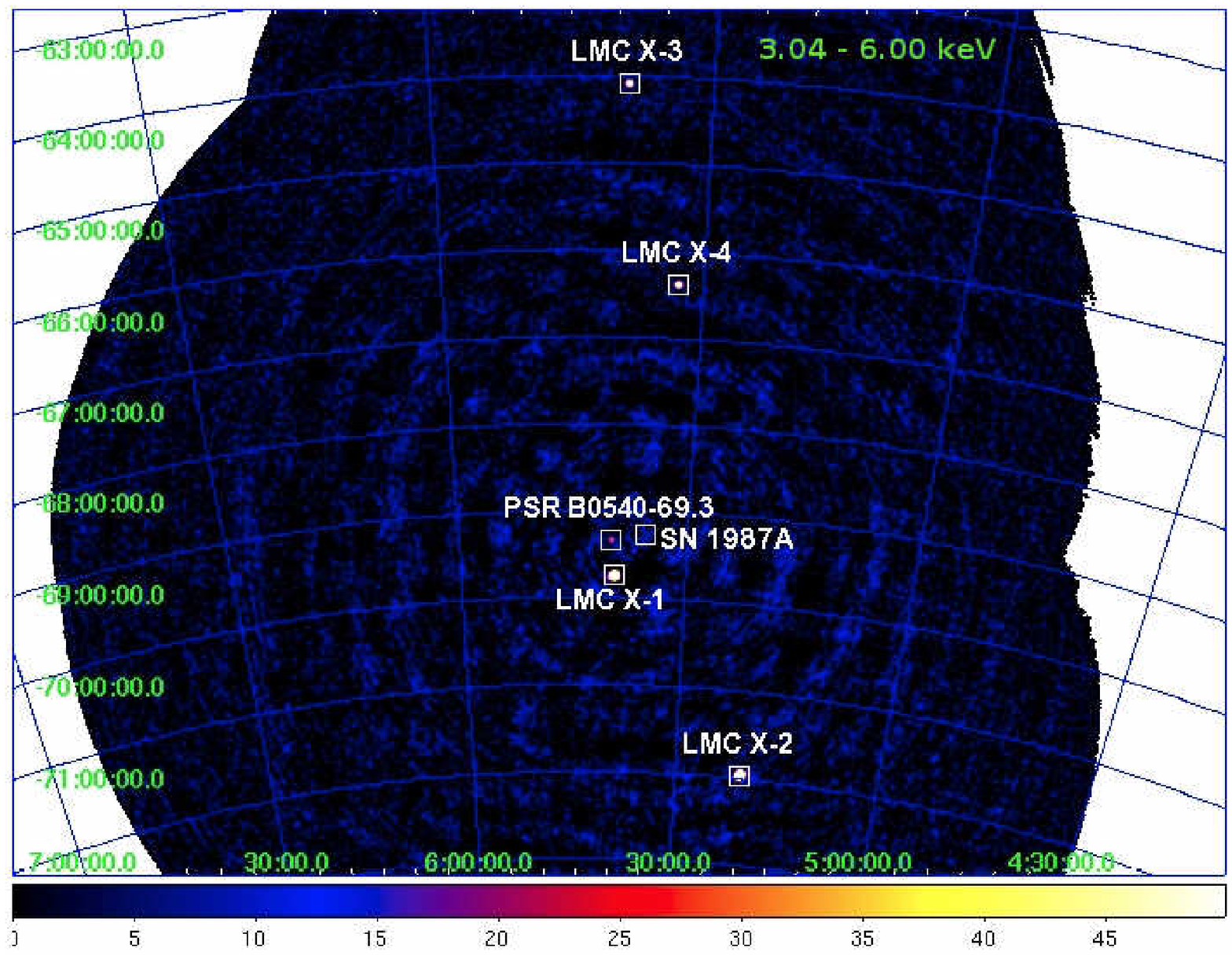,width=0.42\textwidth}
\psfig{file=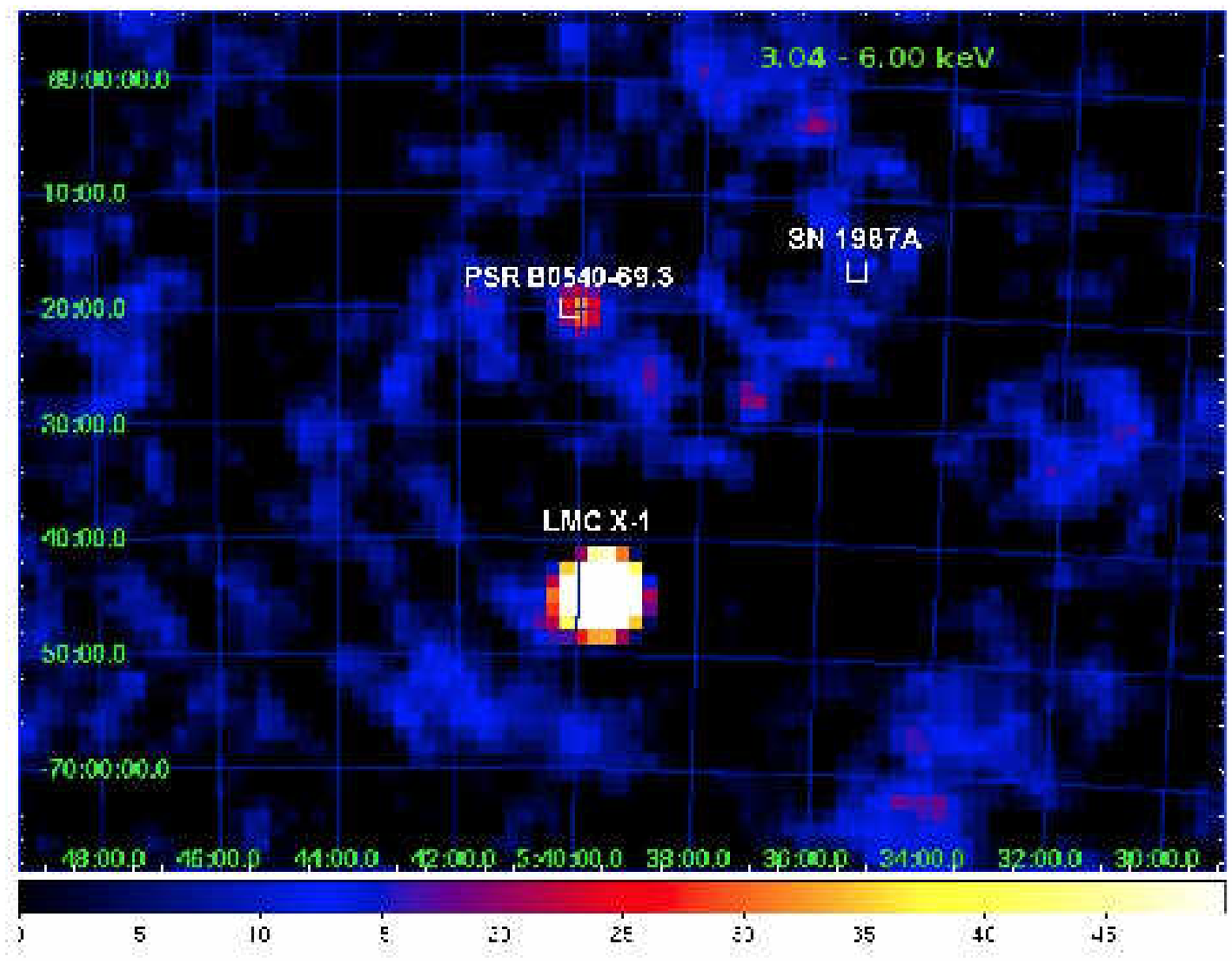,width=0.42\textwidth}
}}
\centerline{\hbox{
\psfig{file=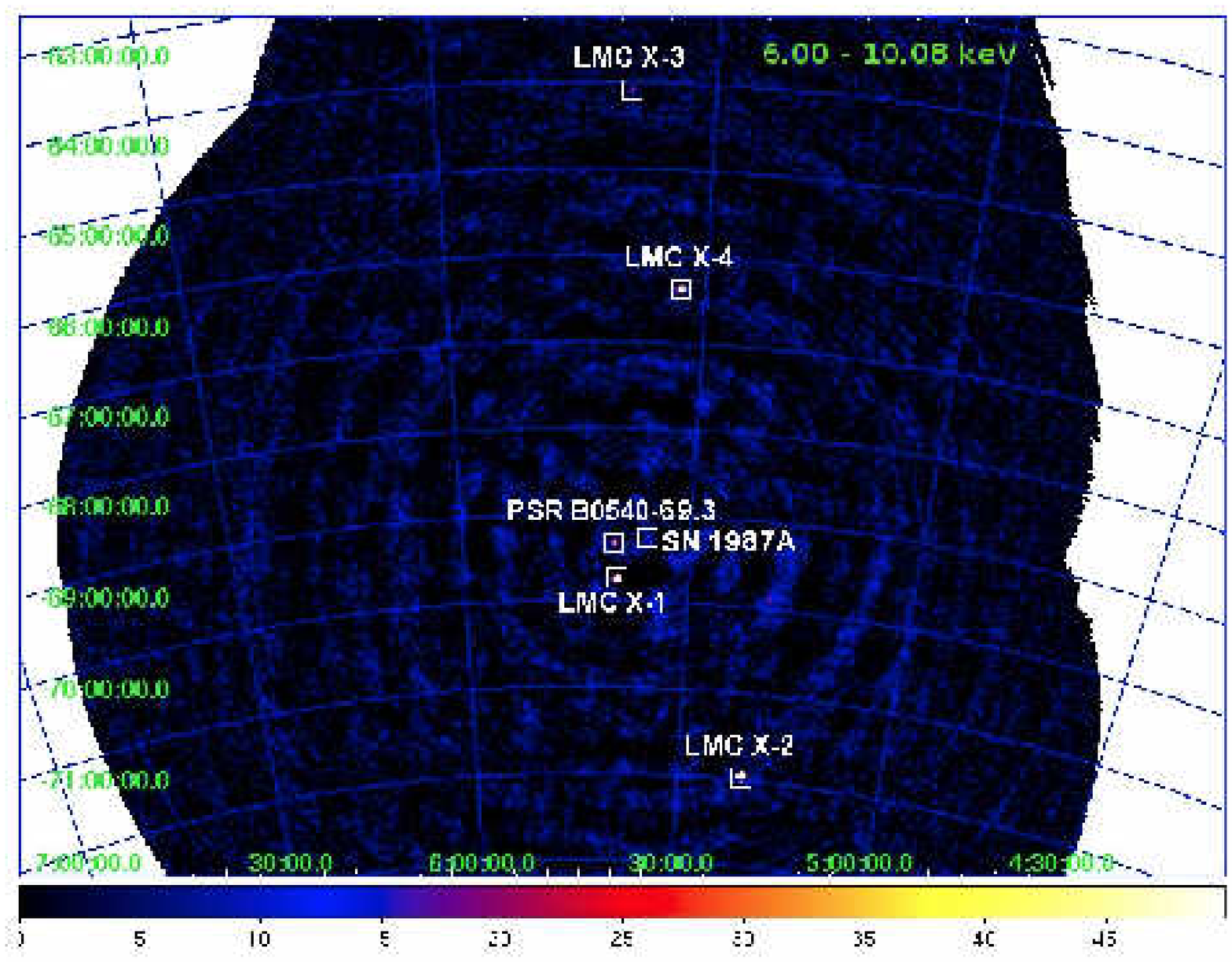,width=0.42\textwidth}
\psfig{file=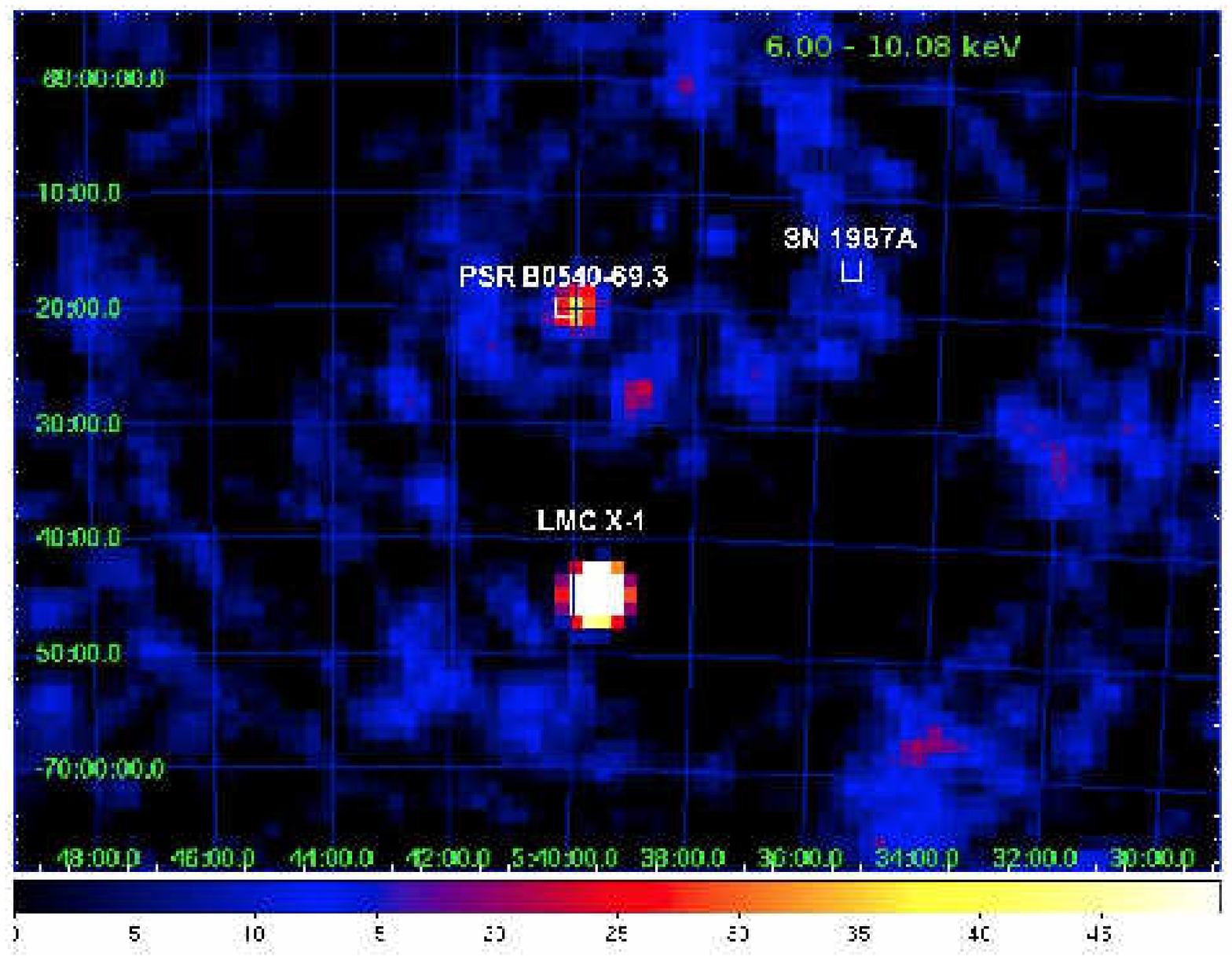,width=0.42\textwidth}
}}
\centerline{\hbox{
\psfig{file=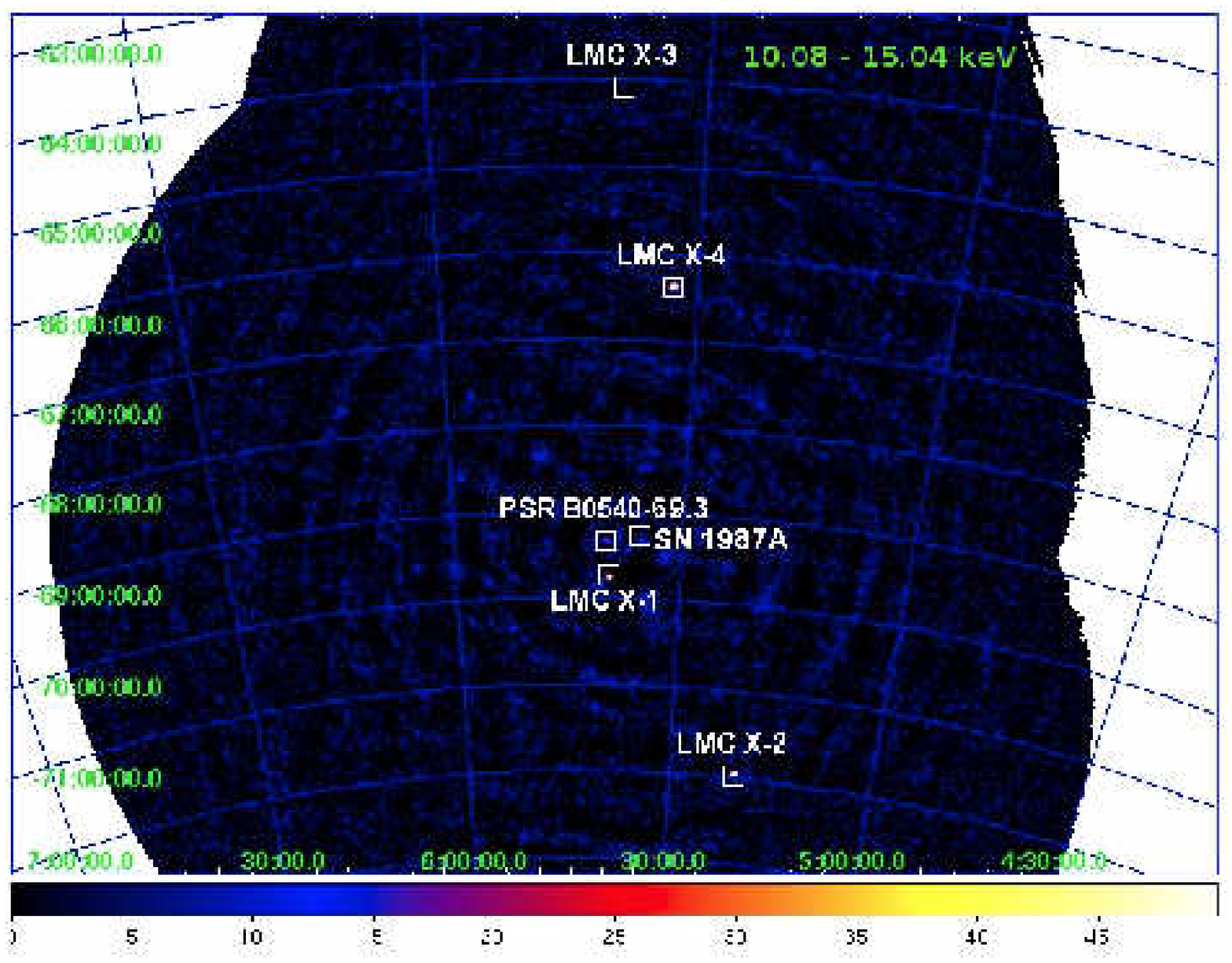,width=0.42\textwidth}
\psfig{file=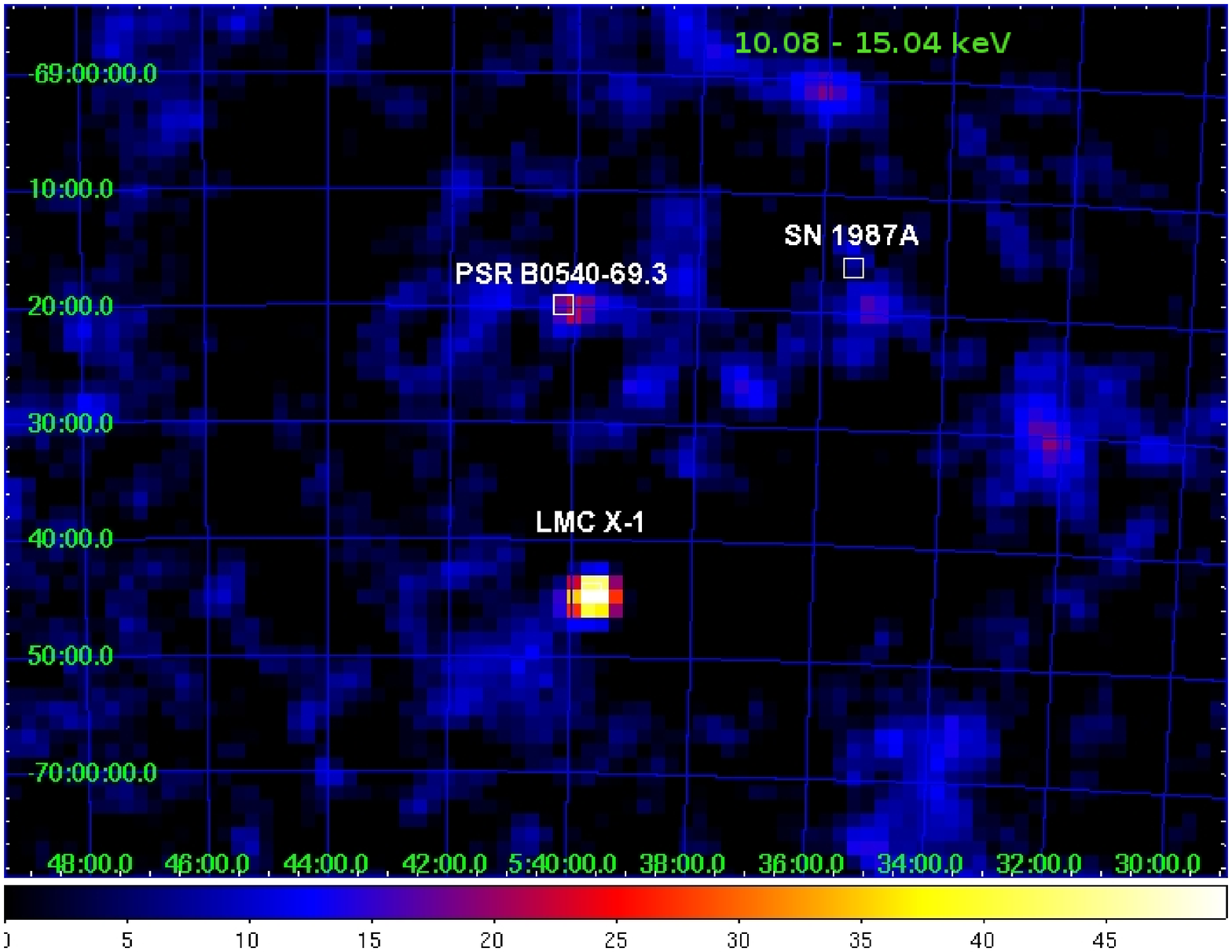,width=0.42\textwidth}
}}
\centerline{\hbox{
\psfig{file=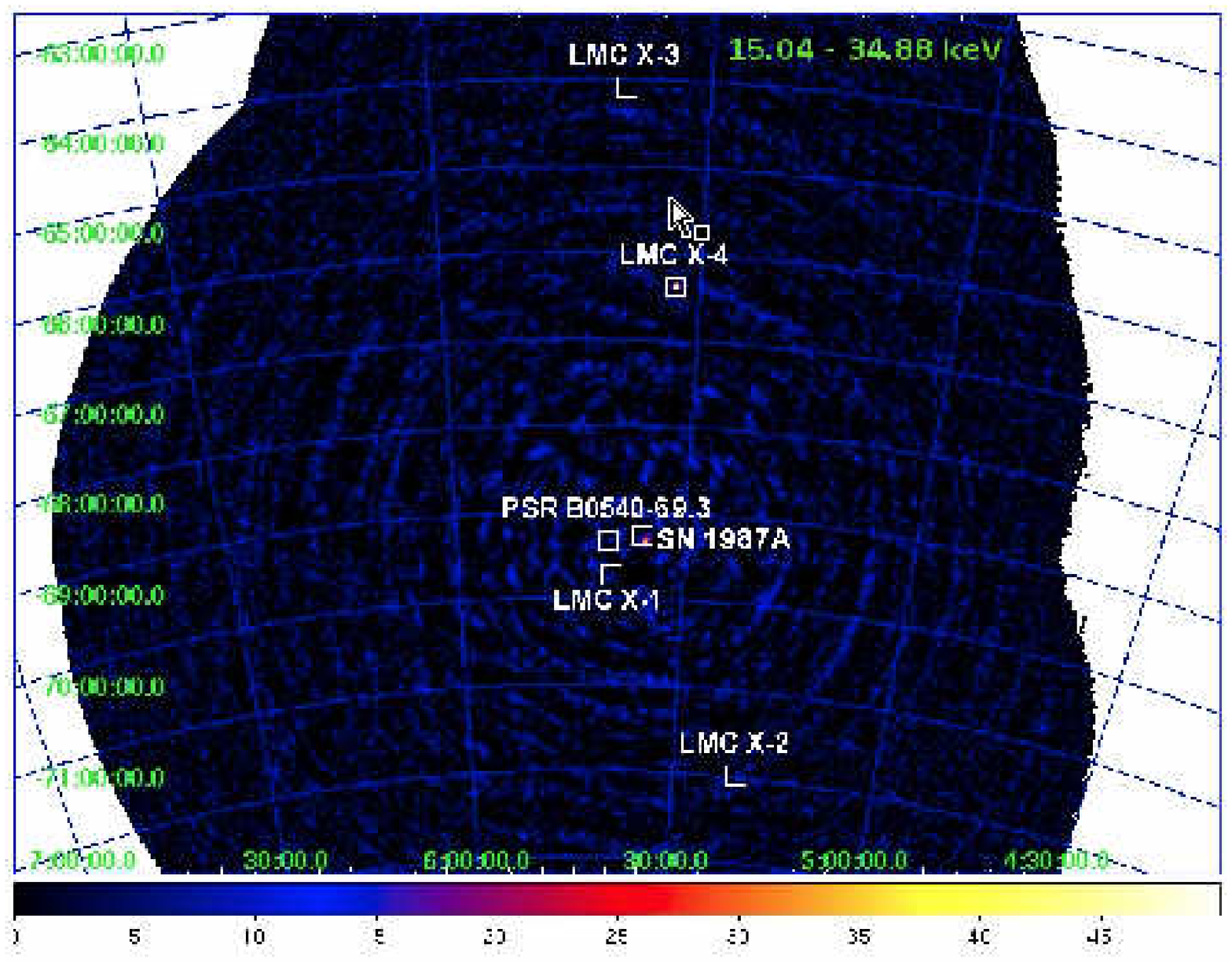,width=0.42\textwidth}
\psfig{file=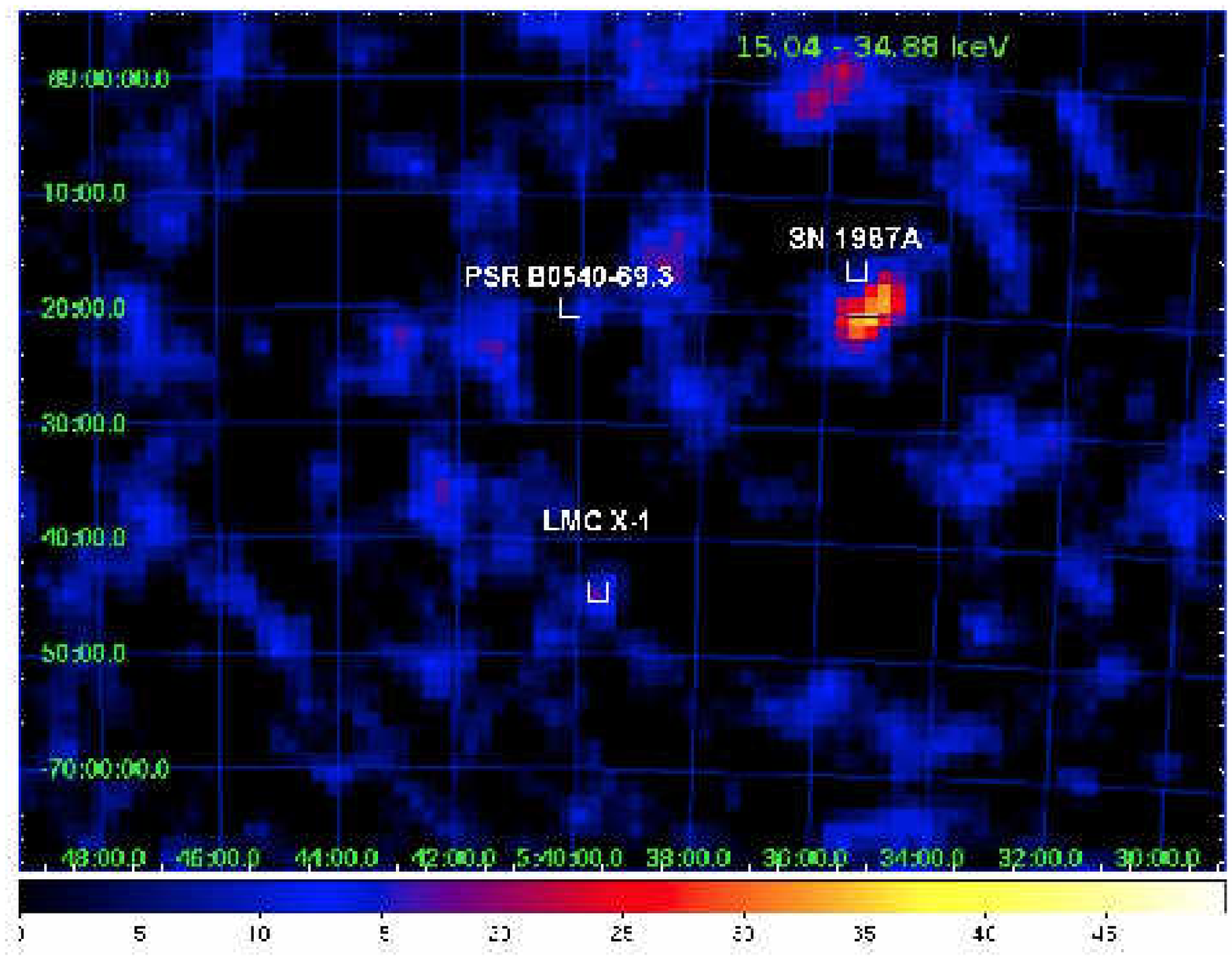,width=0.42\textwidth}
}}
\caption{The JEM-X significance mosaic images of the LMC region (\emph{left}) and the close-up of the pulsar neighbourhood (\emph{right}). Maps are for the 3--6, 6--10, 10--15 and 15--35~keV energy range from top to bottom, respectively.  PSR B0540-69 has been detected with following significance 32.6$\sigma$, 37.1$\sigma$, 21.9$\sigma$ and 11.6$\sigma$, respectively.}
\label{Fig:jemx_mosa2}
\end{figure*}

The Joint European Monitor for X-rays (JEM-X) operates simultaneously with the main $\gamma$-ray \emph{INTEGRAL} instruments. It is based on the same principles as IBIS and SPI instruments: sky imaging accomplished by a coded aperture mask based on a Hexagonal Uniformly Redundant Array (HURA). Its energy range is 3--35~keV, the angular resolution at FWHM is $3'$, the fully illuminated field of view is $4.8\degr$, and the time resolution is on the level of $122~\rm \mu s$ and $1~\rm ms$ for relative and absolute timing, respectively. We used the data of the full imaging telemetry format only, i.e. the image resolution of the detector was 256x256 pixels, timing resolution $1/8192~\rm s = 122~\rm \mu s$, and spectral resolution 256 Pulse Height Amplitude (PHA) channels. The position determination accuracy depends on the number of source and background counts and on the position in the FoV. The off-axis collimator blocks  some of the source photons and beyond the fully coded FoV the coding is incomplete. There has been a shift in the position correction procedure after the beginning of the reprocessing \textit{INTEGRAL} revision 1 data to the revision 2 at ISDC. Therefore, when using \textsf{jemx\_science\_analysis} we always started the analysis from the correction (COR) level, even for Rev.2 data.

At the beginning 409 ScWs fulfilled our selection criteria for JEM-X. After the first run of \textsf{jemx\_science\_analysis} 388 ScWs remained. Some of them showed a `negative selection' error, whereas some did not have the fully codded imaging events. Similarly to the IBIS/ISGRI analysis we used the user catalogue to obtain the images and light curves of PSR B0540-69. The list of sources included during the analysis is the same as in the case of ISGRI, except EXO~0748-676. This source was not in the FoV of JEM-X during the LMC observations.

The source was not detected above the $3 \sigma$ level in almost all single ScWs. Distribution of the detection significance of PSR B0540-69.3 for 388 single JEM-X ScWs gives: mean $1.35$, and standard deviation $0.84$. All the individual images from the different science windows gathered in the observation group can be combined in the second step of image reconstruction, and produce JEM-X mosaic images. The combined images have longer exposure time. As a consequence, weaker sources which are not visible in single ScW can appear in the mosaic images. This is the case of the PSR~B0540-69, Fig.~\ref{Fig:jemx_mosa2}. The JEM-X light curves were obtained in similar way as the IBIS/ISGRI light curves. The pulsar profile in the following energy bands: 3--6, 6--10, 10-15~keV are shown in Fig.~\ref{Fig:lc_0540} (\emph{left column}).

\section{Summary}
The INTEGRAL analysis presented here is based on the \emph{INTEGRAL} observations of the LMC obtained in January 2003 and 2004 with a total exposure of $\sim 1.5$~Ms. In the mosaic maps from the total exposure (JEM-X and IBIS/ISGRI) a source at the location of PSR B0540-69 is clearly visible up to energies of ~200 keV. After barycentric correction and determination of the pulsar phases, based on the ephemeris available from contemporaneous RXTE data, the light curves show the characteristic shape of a broad pulse up into the 40-100 keV band. At higher energies no significant pulsation is detectable. 

For the first time PSR B0540-69 has been detected up to $\sim200$~keV with pulsations visible up to 100 keV. The total source photon spectrum can be fitted with a power law with photon index -2.2 and the flux in the range $17-300$~keV is about $6 \times 10^{-11}~\rm{erg s^{-1} cm^{-2}}$. The pulsed fraction of the total emission decreases with energy and only upper limits could be derived above 100 keV assuming a light curve profile from lower energies. \citet{Cheng1995} predict in an outer gap emission model for the parameters of PSR B0540-69 a significant downturn of the synchrotron spectrum around 50 keV, which seems to be confirmed by our result.  Above 100 keV the \citeauthor{Cheng1995} model predicts an inverse Compton spectrum to dominate, but its intensity would be lower by about a factor of 10 with respect to the extrapolation from soft X-rays and therefore undetectable with the presently available sensitivity.

\begin{acknowledgements}
We thank Lucien Kuiper (SRON, Utrecht) for his help with the RXTE pulsar ephemerides. AS acknowledges the support from the Polish grant 2P03D.004.24. AS also acknowledges support from the Deutscher Akademischer Austausch Dienst (DAAD) grant A/05/15133.
\end{acknowledgements}

\bibliographystyle{aa}
\bibliography{as_lmc.bib}

\begin{thebibliography}{17}
\expandafter\ifx\csname natexlab\endcsname\relax\def\natexlab#1{#1}\fi

\bibitem[{{Boyd} {et~al.}(1995){Boyd}, {van Citters}, {Dolan}, {Wolinski},
  {Percival}, {Bless}, {Elliot}, {Nelson}, \& {Taylor}}]{Boyd1995}
{Boyd}, P.~T., {van Citters}, G.~W., {Dolan}, J.~F., {et~al.} 1995, \apj, 448,
  365

\bibitem[{{Chanan} {et~al.}(1984){Chanan}, {Helfand}, \&
  {Reynolds}}]{Chanan1984}
{Chanan}, G.~A., {Helfand}, D.~J., \& {Reynolds}, S.~P. 1984, \apjl, 287, L23

\bibitem[{{Cheng} \& {Wei}(1995)}]{Cheng1995}
{Cheng}, K.~S. \& {Wei}, D.~M. 1995, \apj, 448, 281

\bibitem[{{de Plaa} {et~al.}(2003){de Plaa}, {Kuiper}, \&
  {Hermsen}}]{dePlaa2003}
{de Plaa}, J., {Kuiper}, L., \& {Hermsen}, W. 2003, \aap, 400, 1013

\bibitem[{{Deeter} {et~al.}(1999){Deeter}, {Nagase}, \& {Boynton}}]{Deeter1999}
{Deeter}, J.~E., {Nagase}, F., \& {Boynton}, P.~E. 1999, \apj, 512, 300

\bibitem[{{Gotthelf} \& {Wang}(2000)}]{Gotthelf2000}
{Gotthelf}, E.~V. \& {Wang}, Q.~D. 2000, \apjl, 532, L117

\bibitem[{{G{\"o}tz} {et~al.}(2006){G{\"o}tz}, {Mereghetti}, {Merlini},
  {Sidoli}, \& {Belloni}}]{Gotz2006}
{G{\"o}tz}, D., {Mereghetti}, S., {Merlini}, D., {Sidoli}, L., \& {Belloni}, T.
  2006, \aap, 448, 873

\bibitem[{{Hirayama} {et~al.}(2002){Hirayama}, {Nagase}, {Endo}, {Kawai}, \&
  {Itoh}}]{Hirayama2002}
{Hirayama}, M., {Nagase}, F., {Endo}, T., {Kawai}, N., \& {Itoh}, M. 2002,
  \mnras, 333, 603

\bibitem[{{Johnston} \& {Romani}(2003)}]{Johnston2003}
{Johnston}, S. \& {Romani}, R.~W. 2003, \apjl, 590, L95

\bibitem[{{Kaaret} {et~al.}(2001){Kaaret}, {Marshall}, {Aldcroft}, {Graessle},
  {Karovska}, {Murray}, {Rots}, {Schulz}, \& {Seward}}]{Kaaret2001}
{Kaaret}, P., {Marshall}, H.~L., {Aldcroft}, T.~L., {et~al.} 2001, \apj, 546,
  1159

\bibitem[{{Manchester} {et~al.}(1993){Manchester}, {Mar}, {Lyne}, {Kaspi}, \&
  {Johnston}}]{Manchester1993}
{Manchester}, R.~N., {Mar}, D.~P., {Lyne}, A.~G., {Kaspi}, V.~M., \&
  {Johnston}, S. 1993, \apjl, 403, L29

\bibitem[{{Middleditch} \& {Pennypacker}(1985)}]{Middleditch1985}
{Middleditch}, J. \& {Pennypacker}, C. 1985, \nat, 313, 659

\bibitem[{{Mineo} {et~al.}(1999){Mineo}, {Cusumano}, {Massaro}, {Nicastro},
  {Parmar}, \& {Sacco}}]{Mineo1999}
{Mineo}, T., {Cusumano}, G., {Massaro}, E., {et~al.} 1999, \aap, 348, 519

\bibitem[{{Serafimovich} {et~al.}(2004){Serafimovich}, {Shibanov}, {Lundqvist},
  \& {Sollerman}}]{Serafimovich2004}
{Serafimovich}, N.~I., {Shibanov}, Y.~A., {Lundqvist}, P., \& {Sollerman}, J.
  2004, \aap, 425, 1041

\bibitem[{{Seward} {et~al.}(1984){Seward}, {Harnden}, \&
  {Helfand}}]{Seward1984}
{Seward}, F.~D., {Harnden}, F.~R., \& {Helfand}, D.~J. 1984, \apjl, 287, L19

\bibitem[{{Taylor} \& {Cordes}(1993)}]{Taylor1993}
{Taylor}, J.~H. \& {Cordes}, J.~M. 1993, \apj, 411, 674

\bibitem[{{Walter} {et~al.}(2003){Walter}, {Rodriguez}, {Foschini}, {de Plaa},
  {Corbel}, {Courvoisier}, {den Hartog}, {Lebrun}, {Parmar}, {Tomsick}, \&
  {Ubertini}}]{Walter2003}
{Walter}, R., {Rodriguez}, J., {Foschini}, L., {et~al.} 2003, \aap, 411, L427

\end{thebibliography}
   
  \clearpage

\end{document}